\documentclass[12pt]{article}

\usepackage{graphicx}
\usepackage{amsmath}
\usepackage{mathptmx}      % use Times fonts if available on your TeX system
\usepackage{enumerate}
\usepackage[misc]{ifsym}
\usepackage{amsmath}
\usepackage{amssymb}
\usepackage{graphicx}
\usepackage{epsf}
\usepackage{epsfig}
\usepackage{color}
\usepackage{array} 
\usepackage{epstopdf}
\newcolumntype{L}{>{\displaystyle}l} 
\newcolumntype{C}{>{\displaystyle}c} 
\newcolumntype{R}{>{\displaystyle}r} 
\usepackage{parskip}

\newcommand{\fatsigma}{\boldsymbol{\sigma}}
\newcommand{\fatvarsigma}{\boldsymbol{\varsigma}}

\newcommand{\fatxi}{\boldsymbol{\xi}}
\newcommand{\fatzeta}{\boldsymbol{\zeta}}

\newcommand{\fatA}{\mathbf{A}}
\newcommand{\fata}{\mathbf{a}}
\newcommand{\fatB}{\mathbf B}
\newcommand{\fatC}{\mathbf C}
\newcommand{\fatc}{\mathbf c}

\newcommand{\fatD}{\mathbf D}
\newcommand{\fate}{\mathbf e}

\newcommand{\fatF}{\mathbf F}
\newcommand{\fatg}{\mathbf g}
\newcommand{\fath}{\mathbf h}
\newcommand{\dfath}{\delta{\mathbf h}}
\newcommand{\fatI}{\mathbf I}
\newcommand{\fatn}{\mathbf n}

\newcommand{\fatt}{\mathbf t}
\newcommand{\fatT}{\mathbf T}
\newcommand{\fatx}{\mathbf x}
\newcommand{\fatu}{\mathbf u}
\newcommand{\dfatu}{\delta\mathbf u}
\newcommand{\fatv}{\mathbf v}
\newcommand{\dfatv}{\delta\mathbf v}

\newcommand{\dpsi}{\delta \psi}

\newcommand{\heta}{\hat{\eta}}
\newcommand{\db}{\delta b}
\renewcommand{\dh}{\delta h}
\newcommand{\dH}{\delta H}
\newcommand{\dC}{\delta C}
\newcommand{\dL}{\delta \calL}
\newcommand{\ddp}{\delta p}
\newcommand{\dq}{\delta q}
\newcommand{\du}{\delta u}

\newcommand{\calL}{\mathcal{L}}

\newcommand{\tfateta}{\tilde{\boldsymbol{\eta}}}
\newcommand{\tfatsigma}{\tilde{\fatsigma}}
\newcommand{\tfatvarsigma}{\tilde{\fatvarsigma}}

\newcommand{\calA}{\mathcal{A}}
\newcommand{\calB}{\mathcal{B}}
\newcommand{\calF}{\mathcal{F}}
\newcommand{\fatzero}{\mathbf{0}}
\newcommand{\tr}{{\rm tr}\,}
\newcommand{\calI}{\mathcal{I}}
\renewcommand{\d}{\text{d}}
\newcommand{\pdd}[2]{\frac{\partial #1}{\partial #2}}

\newcommand{\fatvInv}{{\fatv}}
\newcommand{\qInv}{{q}}
\newcommand{\psiInv}{{\psi}}
\newcommand{\fatxInv}{\bar{\fatx}}

\usepackage[authoryear]{natbib}
\bibpunct{(}{)}{}{a}{}{;}

\begin{document}

\title{Parameter sensitivity analysis of dynamic ice sheet models
\thanks{This work was funded by FORMAS 2017-00665}
}

\author{Cheng Gong, Per L{\"o}tstedt\\
Department of Information Technology. \\  Uppsala University, SE-751 05 Uppsala, Sweden
}

% \institute{G. Cheng (\Letter) \at
%               Division of Scientific Computing, Department of Information Technology, Uppsala University, Uppsala, Sweden \\
%               Tel.: +46-76-2277808\\
%               % Fax: +123-45-678910\\
%               \email{cheng.gong@it.uu.se} %  \\
% %             \emph{Present address:} of F. Author  %  if needed
%            \and
%            P. L{\"o}tstedt \at
%               Division of Scientific Computing, Department of Information Technology, Uppsala University, Uppsala, Sweden\\
%               \email{per.lotstedt@it.uu.se})
% }

% % The correct dates will be entered by the editor

\maketitle

\begin{abstract}
  The velocity field and the height at the surface of a dynamic ice sheet are observed. 
  The ice sheets are modeled by the full Stokes equations and shallow shelf/shelfy stream approximations.
  Time dependence is introduced by a kinematic free surface equation which updates the surface elevation using the velocity solution.
  The sensitivity of the observed quantities at the ice surface to parameters in the models, for example the basal topography and friction coefficients, is analyzed by first deriving the time dependent adjoint equations. 
  Using the adjoint solutions, the effect of a perturbation in a parameter is obtained showing the importance of including the time dependence, in particular when the height is observed.
  The adjoint equations are solved analytically and numerically and the sensitivity of the desired parameters is determined in several examples in two dimensions. 
  A closed form of the analytical solutions to the adjoint equations is given for a two dimensional grounding line migration problem in steady state under the shallow shelf approximation.
  % \keywords{  ice sheet modeling \and inverse problem \and adjoint method \and sensitivity analysis }
\end{abstract}

%%%%%%%%%%%%%%%%%%%%%%%%%%%%%%%%%%%%%%%%%%%%%%%%%%%%%%%%%%%%%%%%%%%%%%%%%%%%%%%%%%%%%%%%%%%%
\section{Introduction}
%%%%%%%%%%%%%%%%%%%%%%%%%%%%%%%%%%%%%%%%%%%%%%%%%%%%%%%%%%%%%%%%%%%%%%%%%%%%%%%%%%%%%%%%%%%%

Large ice sheets cover Antarctica and Greenland, and glaciers are found in mountainous regions all over the world. The ice moves slowly to lower elevations on the bedrock and it floats if it reaches the ocean. There is a need to predict what happens to the ice due to the current and future climate change.

In computational models for the flow of ice in glaciers and continental ice sheets, it is necessary to choose models and equations and to supply parameters for the sliding between the ice base and the bedrock. Direct observations of the basal conditions by drilling holes in the ice are not feasible except for a few locations. 
Instead, the data for the sliding models are inferred from observations of the surface elevation and velocity of the ice from aircraft and satellites, see \citep{Minchew16, Sergienko13}.
How to do this is an important question because the basal sliding is a key uncertainty in the assessment of the future sea level rise due to melting ice \citep{REDPPH15}.

We address here a related question how to determine the sensitivity of the observations at the surface to the conditions at the base. The same solution technique is applicable in sensitivity analysis, uncertainty quantification, and the inverse problem. We extend the adjoint (or control) method by including the time evolution of the ice and its thickness.

The flow of ice is well modeled by the full Stokes (FS) equations, see \citep{GreveBlatterBok}. They form a system of partial differential equations (PDEs) for the stress and pressure in the ice with a nonlinear viscosity coefficient. 
The domain of the ice is confined by an upper surface and a base either resting on the bedrock or floating on sea water.
The boundary conditions at the upper surface of the ice and at the floating part are well defined. 
For the ice in contact with the bedrock, a friction model with parameters determines the sliding force. 
The sliding depends on the topography at the ice base, the friction between the ice and the bedrock, and the meltwater under the ice.
The upper, free boundary of the ice and the interface between the ice and the water are advected by equations for the height and the interface.

The computational effort to solve the FS equations is quite large and there is often a need for approximations. 
The FS equations are simplified by integrating in the depth of the ice in the shallow shelf (or shelfy stream) approximation (SSA) \citep{GreveBlatterBok, SSA}. 
The spatial dimension of the problem is reduced by one with SSA compared to FS.
The pressure is also decoupled from the stress in the system.
% The shallow ice approximation (SIA) is a good approximation in the inner parts of an ice sheet where vertical shear stresses dominate over the longitudinal stresses in the ice flow \citep{GreveBlatterBok, WGH99}. 

The friction model is often of Weertman type \citep{Weertman57} but other models are also considered \citep{tsai2015}.
The model for the relation between the sliding speed and the pressure and the friction at the bed is discussed in \citep{MMPLRGI19, Stearn18}. 
It is not clear how to formulate a relation that is generally applicable. 
When the parameters in the sliding model in the forward equation are unknown in numerical simulations but data are available such as the surface velocity and elevation of the ice, an inverse problem is solved by minimizing the distance between the observations and the predictions of the numerical PDE model with the parameters. 
The gradient of the objective function for the minimization is computed by solving an adjoint equation as in \citep{BrGCGa19, YRSM18}. 
With a fixed thickness of the ice, the adjoint of the FS equations is derived in \citep{Petra12} and for SSA with a frozen viscosity in \citep{MacAyeal2}.

The basal parameters are estimated from uncertain observational data at the surface in \citep{GCDGMMRR16, Isaac15} and initial data for ice sheet simulations are found in \citep{Perego14} using the same technique as in \citep{Petra12}.
The sensitivity of the ice flow to the basal conditions is investigated in \citep{Heimbach12} with the adjoint solution.
Part of the drag at the base may be due to the resolution of the topography. 
The geometry at the ice base is inferred by an inversion method in \citep{vanPelt13}.
The difficulty to separate the topography from the sliding properties at the base in the inversion is also addressed in \citep{KSGF18, TRGBVJ03}. 
Considerable differences in the friction coefficient in the FS and SSA models are found after inversion in \citep{SDEEMGC19}.
By linearization of the model equations, a transfer operator is derived in \citep{Gudmundsson03, Gudmundsson08a} and it is shown in \citep{Gudmundsson08b} how the topography and the friction coefficients are affected by measurement errors at the surface.

It is noted in \citep{Vallot17} that the friction coefficient varies in space and time. 
The time scales of the variations are diurnal \citep{Schoof10}, seasonal \citep{Sole11}, and decennial \citep{JayAllemand11}. 
%Other time dependent forces are considered in \citep{SGSTMS19}. 
%The effect of seasonal variation of the lubrication at the base of the ice is studied in \citep{Shannon13} for the Greenland ice sheet by solving the FS and other high order equations.
The time it takes for the surface to respond to sudden changes in basal conditions are determined analytically and numerically in \citep{Gudmundsson08a} with SSA and FS. 
Time dependent data are used in \citep{Goldberg15} to infer time independent parameters in an ice model.
By applying an inverse method in \citep{JayAllemand11}, the authors observe that a friction parameter varies several orders of magnitude in a decade at the bottom of a glacier.
%Fast temporal variations in the meltwater under the ice drive the ice flow in the analysis in \citep{Schoof10}. 
%The spatial and temporal variations of the basal conditions are inferred from satellite data in \citep{Larour14} with an inverse method for SSA and automatic differentiation.
%Based on observations, the conclusion in \citep{Sole11} is also that the annual change of the water drainage under the ice affects the sliding and the acceleration and deceleration of the ice.
These papers indicate that it is not sufficient to infer the friction parameters from the time-independent adjoint to the FS stress equation but to include also the time dependent height advection equation in the inversion. 

In this paper, we study how perturbations in the sliding conditions and the topography at the base of the ice affect observations of the height of the ice and the velocity at the surface for the FS and SSA models.
The friction law is due to Weertman \citep{Weertman57}.
The sensitivity to perturbations of the parameters in FS and SSA is determined by solving an adjoint problem of the same kind as for inverse problems with a fixed
ice thickness in \citep{MacAyeal2, Petra12}. 
The difference here is that the adjoint of the time dependent advection equation for the height is also solved allowing height observations and the influence is evaluated of the adjoint height in the adjoint solution.
The adjoint equations follow from the Lagrangian of the forward equations after partial integration. The relation is derived between the inverse problem to infer basal parameters from surface data and the sensitivity problem to observe changes in the velocity or the height at the surface from perturbations of the basal parameters.
The stationary adjoint equations are solved analytically in two dimensions under simplifying but reasonable assumptions
making the dependence of the parameters explicit. 
The time dependent adjoint equations are solved numerically for sensitivity analysis of perturbations varying in time.
% The sensitivity in SIA solutions is found algebraically without solving differential equations. 
Conclusions are drawn in the final section. Extensive numerical computations are reported in a companion paper \citep{CGPL19a} illustrating the accuracy of the analytical solutions and the adjoint approach.

%%%%%%%%%%%%%%%%%%%%%%%%%%%%%%%%%%%%%%%%%%%%%%%%%%%%%%%%%%%%%%%%%%%%%%%%%%%%%%%%%%%%%%%%%%%%%%%%%%%%%%%%%%%%%%%%%%%%%%%%%%%%%%%%%%%%%%%%%%%%%%%%%%%%%
\section{Ice models} \label{sec:mathmodel}
%%%%%%%%%%%%%%%%%%%%%%%%%%%%%%%%%%%%%%%%%%%%%%%%%%%%%%%%%%%%%%%%%%%%%%%%%%%%%%%%%%%%%%%%%%%%%%%%%%%%%%%%%%%%%%%%%%%%%%%%%%%%%%%%%%%%%%%%%%%%%%%%%%%%%

The full Stokes equations in glaciology are system of nonlinear PDEs which consists of conservation laws of mass, momentum and energy \citep{GreveBlatterBok}.
% for the flow of ice on the ground and ice floating on water
The nonlinearity is due to the viscosity of the ice according to Glen's flow law \citep{glen1955}.
The ice is assumed to be isothermal and obeys a friction law at the contact surface between the ice and the bedrock. 
The upper surface of the ice is a moving boundary and satisfies an advection equation for the height of the ice. 
The aspect ratio of a continental ice-sheet and a floating ice-shelf, i.e. the thickness scale $V$ divided by the length scale $L$, is low $V/L\sim 10^{-2}$ to $10^{-3}$. 
This scale difference is used in the SSA assumption to simplify the FS equations, see \citep{GreveBlatterBok}.

%%%%%%%%%%%%%%%%%%%%%%%%%%%%%%%%%%%%%%%%%%%%%%%%%%%%%%%%%%%%%%%%%%%%%%%%%%%%%%%
%%%%%%% Full Stokes
%%%%%%%%%%%%%%%%%%%%%%%%%%%%%%%%%%%%%%%%%%%%%%%%%%%%%%%%%%%%%%%%%%%%%%%%%%%%%%%
\subsection{Full Stokes equation} \label{sec:FS}

%%%%%%%%%%%%%%%%%%%%%%%%%%%
% Ice geometry
%%%%%%%%%%%%%%%%%%%%%%%%%%%
\begin{figure}[htbp]
  \begin{center}
    \includegraphics[width=0.50\linewidth]{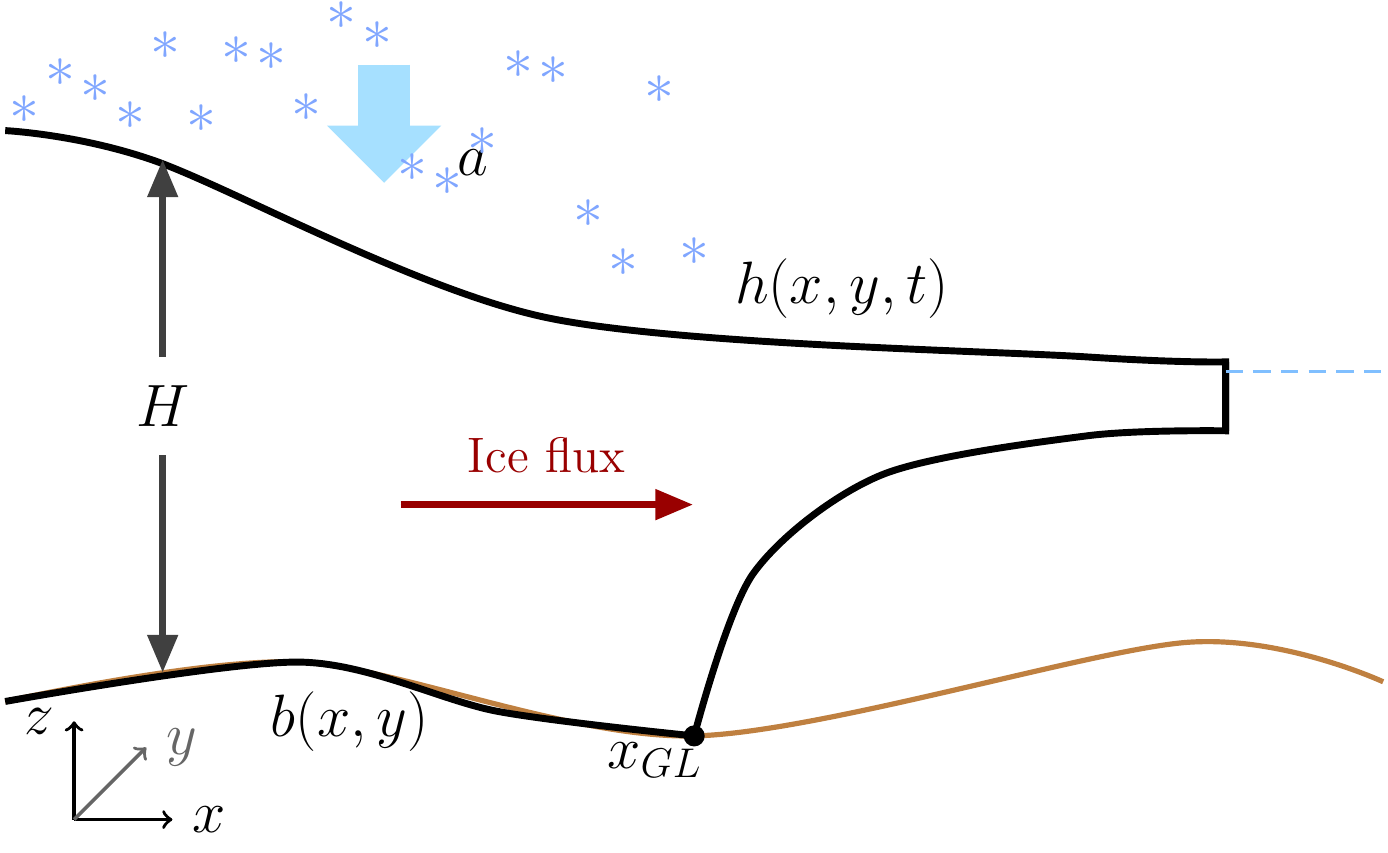}
    \includegraphics[width=0.48\linewidth]{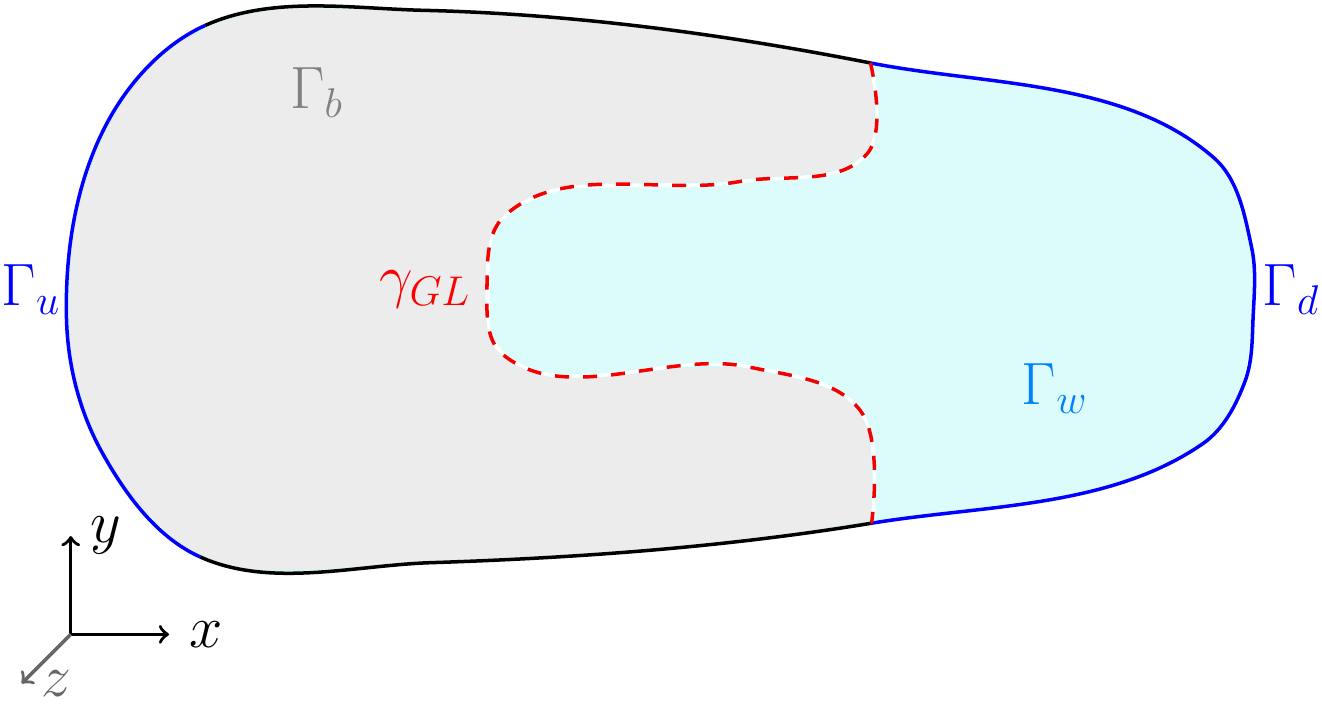}
  \end{center}
  \caption{A schematic view of an ice sheet in the $x-z$ (left panel) and $x-y$ (right panel)  plane. 
        % The grounding line is at $x=x_{GL}$.
        % Left: The upper surface $\Gamma_s$ is at $z=h(x, y, t)$, the grounded boundary $\Gamma_b$ is at $z=z_b(x, y, t)=b(x,y)$ for $x<x_{GL}$, and the floating boundary $\Gamma_w$ at $z=z_b(x, y, t)$ for $x>x_{GL}$.
        % Right: The grounded part of the base is $\Gamma_b$ and the floating part is $\Gamma_w$ separated by the grounding line $\gamma_{GL}$.
           }
  \label{fig:Icegeometry}
\end{figure}

%%%%%%%%%%%%%%%%%%%%%%%%%%%
% \begin{figure}[htbp]
%   \begin{center}
%   \end{center}
%   \caption{A schematic view of an ice sheet in the $x-y$ plane. },
%   \label{fig:3D omega}
% \end{figure}
%%%%%%%%%%%%%%%%%%%%%%%%%%%

Let $u_1, u_2,$ and $u_3$ be the velocity components of $\fatu=(u_1,u_2,u_3)^T$ in the $x, y$ and $z$ directions and $\fatx=(x, y, z)^T$ in three dimensions. 
Vectors and matrices are written in bold characters. 
The horizontal plane is spanned by the $x$ and $y$ coordinates and $z$ is the coordinate in the vertical direction.
Let the subscript $x, y, z$ or $t$ denote a derivative with respect to the variable.
The height of the upper surface is $h(x, y, t)$, the $z$ coordinates of the bedrock and the floating interface are $b(x, y)$ and $z_b(x, y, t)$, and the ice thickness is $H=h-b$, as shown in Fig.~\ref{fig:Icegeometry}.

The strain rate $\fatD$  and the viscosity $\eta$ are given by 
\begin{equation}
\begin{array}{rll}
\label{eq:Ddef}
   \fatD(\fatu)=\frac{1}{2}(\nabla\fatu+\nabla\fatu^T),\; \eta(\fatu)=\frac{1}{2}{A}^{-\frac{1}{n}}({\rm tr}\fatD^2(\fatu))^{\nu},\; \nu=\frac{1-n}{2n},
\end{array}
\end{equation}
where ${\rm tr}\fatD^2$ is the trace of $\fatD^2$. 
The rate factor ${A}$ in \eqref{eq:Ddef} depends on the temperature and Glen's flow law determines $n>0$, here taken to be $n=3$.
The stress tensor is 
\begin{equation}
\label{eq:stress}
   \fatsigma(\fatu, p)=2\eta\fatD(\fatu)-\fatI p,
\end{equation}
where $p$ is the pressure and $\fatI$ is the identity matrix.
The domain occupied by the ice is $\Omega$ with boundary $\Gamma$ whose outward pointing normal is $\fatn$. 
If $\fatx\in\Omega$ then $(x, y)\in\omega$ in the $x-y$ plane. 
The upper boundary is $\Gamma_s$ at the ice surface.
The vertical, lateral boundary has an upstream part $\Gamma_u$ where $\fatn\cdot\fatu\le 0$ and a downstream part $\Gamma_d$ where $\fatn\cdot\fatu> 0$. 
The lower boundary at the bedrock is $\Gamma_b$ and the floating boundary is $\Gamma_w$. 
They are separated by the grounding line $\gamma_{GL}$ defined by $(x_{GL}(y), y)$ by assuming that the ice mainly flows along the $x$-axis.
If $\fatx\in\Gamma_u$ or $\fatx\in\Gamma_d$ then $(x, y)\in\gamma_u$ or $(x, y)\in\gamma_d$ where 
$\gamma=\gamma_u\cup\gamma_d$ is the boundary of $\omega$. The definitions of these domains are
\begin{equation}\label{eq:Omdef}
\begin{array}{rll}
  \Omega&=\{\fatx| (x,y)\in\omega,\, b(x, y)\le z\le h(x,y, t)\},\\ 
  \Gamma_s&=\{\fatx| (x,y)\in\omega,\, z=h(x,y, t)\},\\ 
  \Gamma_b&=\{\fatx| (x,y)\in\omega,\, z=b(x,y),\, x<x_{GL}(y)\},\\
  \Gamma_w&=\{\fatx| (x,y)\in\omega,\, z=z_b(x,y,t),\, x>x_{GL}(y)\},\\
  \Gamma_u&=\{\fatx| (x,y)\in\gamma_u,\, b(x,y)\le z\le h(x,y,t) \},\\ 
  \Gamma_d&=\{\fatx| (x,y)\in\gamma_d,\, b(x,y)\le z\le h(x,y,t)\},
\end{array}
\end{equation}
and a schematic view in the $x-y$ plane is shown in the right panel of  Fig.~\ref{fig:Icegeometry}.

In two vertical dimensions as in the left panel of Fig.~\ref{fig:Icegeometry}, $\fatx=(x,z)^T,\; \omega=[0, L],\; \gamma_u=0,$ and $\gamma_d=L$ where $L$ is the horizontal length of the domain.

%%%%%%%%%%%%%%%%%%%%%%%%%%%%%%%%%%%%%%%%%%%%%%%%%%%%%%%%%%%%
%  Friction law
%%%%%%%%%%%%%%%%%%%%%%%%%%%%%%%%%%%%%%%%%%%%%%%%%%%%%%%%%%%%
The basal stress on $\Gamma_b$ is related to the basal velocity using an empirical friction law.  
The friction coefficient has a general form $\beta(\fatu, \fatx, t)=C(\fatx, t)f(\fatu)$ where the coefficient $C(\fatx, t)$ is independent of the velocity $\fatu$ and $f(\fatu)$ represents some linear or nonlinear function of $\fatu$. For instance,  $f(\fatu)=\|\fatu\|^{m-1}$ with the norm $\|\fatu\|=(\fatu\cdot\fatu)^{1/2}$ introduces a Weertman type friction law \citep{Weertman57} on $\omega$ with a Weertman friction coefficient $C(\fatx, t)>0$ and an exponent parameter $m>0$. Common choices of $m$ are $1/3$ and $1$.

The density of the ice is denoted by $\rho$,  the accumulation/ablation rate on $\Gamma_s$ by $a$, and the gravitational acceleration by $\fatg$.
A projection \citep{Petra12} on the tangential plane of $\Gamma_b$ is denoted by $\fatT=\fatI-\fatn\otimes\fatn$ where
the Kronecker outer product between two vectors $\fata$ and $\fatc$ or two matrices $\fatA$ and $\fatC$ is defined by
\begin{equation}\label{eq:Krondef}
    (\fata\otimes\fatc)_{ij}=a_i c_j,\quad (\fatA\otimes\fatC)_{ijkl}=A_{ij}C_{kl}.
\end{equation}

With $\fath=(h_x, h_y, -1)^T$ (in two dimensions $\fath=(h_x, -1)^T$),  the forward FS equations for the height and velocity are
\begin{equation}
\begin{array}{rll}\label{eq:FSforw}
   &h_t+\fath\cdot\fatu=a,\; \; {\rm on}\; \Gamma_s,\\
   &h(\fatx, 0)=h_0(\fatx),\; \fatx\in\omega,\quad  h(\fatx,t)=h_\gamma(\fatx, t),\; \fatx\in\gamma_u,\\
   &-\nabla\cdot\fatsigma(\fatu, p)=-\nabla\cdot(2\eta(\fatu)\fatD(\fatu))+\nabla p=\rho \fatg,\quad
     \nabla\cdot\fatu=0, \;{\rm in}\;\Omega(t),\\
   & \fatsigma\fatn=\fatzero, \;{\rm on}\; \Gamma_s,\\
   & \fatT\fatsigma\fatn=-Cf(\fatT\fatu)\fatT\fatu,\quad \fatn\cdot\fatu=0,\;{\rm on}\; \Gamma_b,
\end{array}
\end{equation}
where $h_0(\fatx)$ is the initial height and $h_\gamma(\fatx, t)$ is a given height on the inflow boundary.
The boundary conditions of the velocity on $\Gamma_u$ and $\Gamma_d$ are of Dirichlet type such that 
\begin{equation}
  \label{eq:FSupdown}
  \fatu|_{\Gamma_u}=\fatu_u,\quad
  \fatu|_{\Gamma_d}=\fatu_d,
\end{equation}
where $\fatu_u$ and $\fatu_d$ are some known velocities. 
In a special case where $\Gamma_u$ is at the ice divide, the horizontal velocity is set to $\fatu|_{\Gamma_u}=\fatzero$, and the vertical component of $\fatsigma\fatn$ also vanishes on $\Gamma_u$.

%%%%%%%%%%%%%%%%%%%%%%%%%%%%%%%%%%%%%%%%%%%%%%%%%%%%%%%%%%%%%%%%%%%%%%%%%%%%%%%
%%%%%%% SSA
%%%%%%%%%%%%%%%%%%%%%%%%%%%%%%%%%%%%%%%%%%%%%%%%%%%%%%%%%%%%%%%%%%%%%%%%%%%%%%%
\subsection{Shallow shelf approximation} \label{sec:SSA}

On the ice shelf and the fast flowing region, the basal shear stress is negligibly small and the horizontal velocity is almost constant in the $z$ direction \citep{GreveBlatterBok, SSA, Schoof07}. 
The three dimensional FS problem \eqref{eq:FSforw} on $\Omega$ can be simplified to a two dimensional, horizontal problem with  $\fatx=(x,y)\in\omega$ by SSA, where only the horizontal velocity components $\fatu=(u_1,u_2)^T$ are considered. 
The viscosity for the SSA is 
\begin{equation}
  \label{eq:SSA3Dvisco}
  \eta(\fatu)=\frac{1}{2}A^{-\frac{1}{n}}\left(u_{1x}^2+u_{2y}^2+\frac{1}{4}(u_{1y}+u_{2x})^2+u_{1x}u_{2y}\right)^\nu
             =\frac{1}{2}A^{-\frac{1}{n}}\left(\frac{1}{2}\fatB:\fatD\right)^\nu,
\end{equation}
where $\fatB(\fatu)=\fatD(\fatu)+\nabla\cdot\fatu\, \fatI$ with $\nabla\cdot\fatu=\tr \fatD(\fatu)$. 
The Frobenius inner product between two matrices $\fatA$ and $\fatC$ 
is defined by
\begin{equation}\label{eq:Frobdef}
    \fatA:\fatC=\sum_{ij}A_{ij}C_{ij}.
\end{equation}
The vertically integrated stress tensor $\fatvarsigma(\fatu)$ is given by 
\begin{equation}
\label{eq:SSAdef3D}
  \fatvarsigma(\fatu)=2H\eta\fatB(\fatu),
\end{equation}

Let $\fatn$ be the outward normal vector of the boundary $\gamma=\gamma_u\cup\gamma_d$ and $\fatt$ the tangential vector such that $\fatn\cdot\fatt=0$. 
The friction law is defined as in the FS case where the basal velocity is replaced by the horizontal velocity since the vertical variation is neglected in SSA. 
An example is Weertman's law defined by $\beta(\fatu, \fatx, t)=C(\fatx,t)f(\fatu) = C(\fatx,t)\|\fatu\|^{m-1}$ with a friction coefficient $C(\fatx, t) \ge0$.
In Fig.~\ref{fig:Icegeometry}, $\omega=\Gamma_b\cup\Gamma_w$ and $\gamma_u=\Gamma_u,\, \gamma_d=\Gamma_d$.
% Under the floating ice shelf, $C=0$. 

The ice dynamics system is 
\begin{equation}
\begin{array}{rll}\label{eq:SSAforw3D}
   &h_t+\nabla\cdot(\fatu H)=a,\; \; 0\le t \le T,\; \fatx\in\omega,\\
   &h(\fatx, 0)=h_0(\fatx),\; \fatx\in\omega,\quad h(\fatx,t)=h_\gamma(\fatx,t),\,\fatx\in\gamma_u,\\
   &\nabla\cdot\fatvarsigma-Cf(\fatu)\fatu=\rho g H \nabla h, \;\fatx\in\omega,\\
   &\fatn\cdot\fatu(\fatx,t)=u_{u}(\fatx,t), \fatx\in\gamma_u, \quad \fatn\cdot\fatu(\fatx, t)=u_{d}(\fatx,t), \fatx\in\gamma_d,\\
   &\fatt\cdot\fatvarsigma\fatn=- C_{\gamma}f_{\gamma}(\fatt\cdot\fatu)\fatt\cdot\fatu,\; \fatx\in\gamma_g, \quad \fatt\cdot\fatvarsigma\fatn=0,\; \fatx\in\gamma_w.
\end{array}
\end{equation}
The inflow and outflow normal velocities $u_{u}\le 0$ and $u_{d}> 0$ are specified on $\gamma_u$ and $\gamma_d$.
The lateral side of the ice $\gamma$ is split into $\gamma_g$ and $\gamma_w$ with $\gamma=\gamma_g\cup\gamma_w$. 
There is friction in the tangential direction on $\gamma_g$ which depends on the tangential velocity $\fatt\cdot\fatu$ with the friction coefficient $C_\gamma$ and friction function $f_\gamma$. There is no friction on $\gamma_w$.
The structure of the SSA system \eqref{eq:SSAforw3D} is similar to the FS equations in \eqref{eq:FSforw}. 
However, the velocity $\fatu$ is not divergence free in SSA and $\fatB\ne\fatD$ due to the cryostatic approximation of $p$.

In the case where an ice shelf or a grounding line exists, the floating ice is assumed to be at hydrostatic equilibrium in the seawater. 
A calving front boundary condition  \citep{Schoof07,vanderVeen96} is applied at $\gamma_d$ by the depth integrated stress balance
\begin{equation}
  \label{eq:SSAcf}
  \fatvarsigma(\fatu)\cdot \fatn =\frac{1}{2}\rho g H^2\left(1-\frac{\rho}{\rho_w}\right) \fatn, \;\fatx\in\gamma_d,
\end{equation}
where $\rho_w$ is the density of seawater. With this boundary condition, a calving rate $u_c$ can be determined at the ice front.

\section{Adjoint equations}\label{sec:inv}

We wish to determine the sensitivity of a functional
\begin{equation}
  \label{eq:Fdef}
  \calF=\int_0^T\int_{\Gamma_s} F(\fatu, h)\,\d\fatx\, \d t
\end{equation}
at $\Gamma_s$ in the time interval $[0, T]$ to perturbations in the friction coefficient $C(\fatx, t)$ at the base of the ice and the topography $b(\fatx)$ when $\fatu$ and $h$ satisfy the FS equations \eqref{eq:FSforw} or the SSA equations \eqref{eq:SSAforw3D}.
We introduce a Lagrangian $\calL(\fatu, p, h; \fatv, q, \psi; b, C)$ for a given observation $\calF$ with the forward solution $(\fatu, p, h)$ to \eqref{eq:FSforw} or $(\fatu, h)$ to \eqref{eq:SSAforw3D} and the corresponding adjoint solutions $(\fatv, q, \psi)$ or $(\fatv, \psi)$. 
The adjoint solutions solve the adjoint equations to the FS and SSA equations. 
These equations will be derived using the Lagrangian in this section and Appendix~\ref{sec:appA}.

The effect of the perturbations $\dC$ and $\db$ in $C$ and $b$ on $\calF$
is given by the perturbation $\delta\calL$ in the Lagrangian
\[
\begin{array}{lll}
  \delta\calF&=\delta\calL\\
       &=\calL(\fatu+\dfatu, p+\ddp, h+\dh; \fatv+\dfatv, q+\dq, \psi+\dpsi; b+\db, C+\dC)\\
                   &\quad-\calL(\fatu, p, h; \fatv, q, \psi; b, C). 
\end{array}
\]

Examples of $F(\fatu, h)$ in \eqref{eq:Fdef} are $\|\fatu-\fatu_{\rm obs}\|^2$, $|h-h_{\rm obs}|^2$ in an inverse problem to find $b$ and $C$ to match the observed data $\fatu_{\rm obs}$ and $h_{\rm obs}$ at the surface $\Gamma_s$ as in \citep{GCDGMMRR16, Isaac15, Morlighem13, Petra12},  or $F(\fatu, h)=\frac{1}{T}u_1(\fatx, t)\delta(\fatx-\fatx_\ast)$ with the Dirac delta at $\fatx_\ast$ to  measure the time averaged deviation of the horizontal velocity $u_1$  at $\fatx_\ast$ on the ice surface $\Gamma_s$ with
\[
    \calF=\int_0^T\int_{\Gamma_s} F(\fatu, h)\,\d\fatx\,\d t=\frac{1}{T}\int_0^T u_1(\fatx_\ast, t)\, \d t,
\]
where $T$ is the duration of the observation at $\Gamma_s$.

%%%%%%%%%%%%%%%%%%%%%%%%%%%%%%%%%%%%%%%%%%%%%%%%%%%%%%%%%%%%%%%%%%%%%%%%%%%%%%%
%%%%%%% Adjoint FS
%%%%%%%%%%%%%%%%%%%%%%%%%%%%%%%%%%%%%%%%%%%%%%%%%%%%%%%%%%%%%%%%%%%%%%%%%%%%%
\subsection{Full Stokes equation} \label{sec:invFS}

The definition of the Lagrangian $\calL$ for the FS equations is found in \eqref{eq:FSLag} in Appendix~\ref{sec:appA} where $(\fatv, q, \psi)$ are the Lagrange multipliers corresponding to the forward equations for $(\fatu, p, h)$. 
In order to determine $(\fatv, q, \psi)$, the so-called adjoint problem is solved 
\begin{equation}
\begin{array}{rll}\label{eq:FSadj}
   &\psi_t+\nabla\cdot(\fatu \psi)-\fath\cdot\fatu_z\psi=F_h+F_\fatu\cdot\fatu_z,\; \; {\rm on}\; \Gamma_s,\\
   &\psi(\fatx, T)=0,\; \psi(\fatx, t)=0, \;{\rm on}\; \Gamma_d,\\
   &-\nabla\cdot\tfatsigma(\fatv, q)=-\nabla\cdot(2\tfateta(\fatu)\star\fatD(\fatv))+\nabla q=\fatzero,\quad  \nabla\cdot\fatv=0, \;{\rm in}\;\Omega(t), \\
   & \tfatsigma(\fatv, q)\fatn=-(F_{\fatu}+\psi\fath), \;{\rm on}\; \Gamma_s,\\
   & \fatT\tfatsigma(\fatv, q)\fatn=-Cf(\fatT\fatu)\left(\fatI+\fatF_b(\fatT\fatu)\right)\fatT\fatv,\;{\rm on}\; \Gamma_b,\\
   & \fatn\cdot\fatv=0,\;{\rm on}\; \Gamma_b,
\end{array}
\end{equation}
where the derivatives of $F$ with respect to $\fatu$ and $h$ are 
\begin{equation}
    F_\fatu=\left(\pdd{F}{u_1}, \pdd{F}{u_2}, \pdd{F}{u_3} \right)^T, \quad F_h=\frac{\partial F}{\partial h}.\nonumber
\end{equation}
The adjoint viscosity and adjoint stress \citep{Petra12} are
\begin{equation}
\begin{array}{rll}\label{eq:FSviscadj}
   \tfateta(\fatu)&=\eta(\fatu)\left(\calI+\frac{1-n}{n \fatD(\fatu):\fatD(\fatu)}\fatD(\fatu)\otimes\fatD(\fatu)\right),\\ 
   \tfatsigma(\fatv, q)&=2\tfateta(\fatu)\star\fatD(\fatv)-q\fatI.\\
\end{array}
\end{equation}
The tensor $\calI$ has four indices $ijkl$ and $\calI_{ijkl}=1$ only when $i=j=k=l$, otherwise $\calI_{ijkl}=0$.
In general, $\fatF_b(\fatT\fatu)$ in \eqref{eq:FSadj} is a linearization of the friction law relation $f(\fatT\fatu)$ in \eqref{eq:FSforw} with respect to the variable $\fatT\fatu$. For instance, with a Weertman type friction law, $f(\fatT\fatu)=\|\fatT\fatu\|^{m-1}$, it is
\begin{equation}
     \fatF_b(\fatT\fatu)=\frac{m-1}{\fatT\fatu\cdot\fatT\fatu}(\fatT\fatu)\otimes(\fatT\fatu).
\end{equation}

The $\star$ operation in \eqref{eq:FSviscadj} between a four index tensor $\calA$ and a two index tensor or matrix $\fatC$ is defined by
\begin{equation}\label{eq:stardef}
    (\calA\star\fatC)_{ij}=\sum_{kl}\calA_{ijkl}C_{kl}.
\end{equation}
The perturbation of the Lagrangian function with respect to a perturbation $\dC$ in the slip coefficient $C(\fatx,t)$ is 
\begin{equation}\label{eq:FSgrad}
   \delta\calF=\delta\calL=\int_0^T\int_{\Gamma_b} f(\fatT\fatu)\fatT\fatu\cdot\fatT\fatv\,\, \delta C \,\,\d\fatx\, \d t
\end{equation}
involving the tangential components of the forward and adjoint velocities $\fatT\fatu$ and $\fatT\fatv$ at the ice base $\Gamma_b$.

The detailed derivation of the adjoint equations \eqref{eq:FSadj} and the perturbation of the Lagrangian function \eqref{eq:FSgrad} are given in Appendix~\ref{sec:appA}.3 from the weak form of the FS equations \eqref{eq:FSforw} on $\Omega$, integration by parts, and by applying the boundary conditions as in \citep{Martin14, Petra12}.
The adjoint equations consist of the equations for the adjoint height $\psi$, the adjoint velocity $\fatv$, and the adjoint pressure $q$. 
Compared to the steady state adjoint equation for the FS equation in \citep{Petra12}, an advection equation is added in \eqref{eq:FSadj} for the Lagrange multiplier $\psi(\fatx, t)$ on $\Gamma_s$ with a right hand side depending on the observation function $F$ and one term depending on $\psi$ in the boundary condition on $\Gamma_s$.
The adjoint height equation of $\psi$ can be solved independently of the adjoint stress equation since it is independent of $\fatv$.
If $h$ is observed then the adjoint height equation must be solved together with the adjoint stress equation. Otherwise, the term $\psi\fath$ vanishes in the right hand side of 
the adjoint stress equation and the solution is $\fatv=\fatzero$ with $\delta\calF=0$ in \eqref{eq:FSgrad}.
% The multiplier $\psi$ is non-zero if the right hand side of the advection equation is non-zero. 
% The domain $\Omega$ is time-dependent because the upper surface is governed by $h(\fatx, t)$ from the forward equation. 

%$\psi\fath$ is a ``force'' constraining $\fatv$ on $\Gamma_s$ to satisfy the advection equation there, without this term there is no guarantee that $\fatu$ and $h$ satisfy the height advection equation, 

\subsubsection{Time-dependent perturbations}\label{sec:timedepFS}

Suppose that $u_{1\ast} = u_1(\fatx_\ast, t_\ast)$ is observed at $(\fatx_\ast, t_\ast)$ at the ice surface and that $t_\ast<T$, then 
\[
         u_1(\fatx_\ast, t_\ast)=\calF=\int_0^T\int_{\Gamma_s} F(\fatu)\, \d\fatx\, \d t,
\]
with
\[
        F(\fatu)=u_1\delta(\fatx-\fatx_\ast)\delta(t-t_\ast),\; F_{u_1}=\delta(\fatx-\fatx_\ast)\delta(t-t_\ast),\; F_{u_2}=F_{u_3}=0, \; F_h=0.
\]
The procedure to determine the sensitivity is as follows. 
First, the forward equation \eqref{eq:FSforw} is solved for $\fatu(\fatx, t)$ from $t=0$ to $t=T$. 
Then, the adjoint equation \eqref{eq:FSadj} is solved backward in time for $t\le T$ with $\psi(\fatx, T)=0$ as the final condition. 
Obviously, the solution for $t_\ast<t\le T$ is $\psi(\fatx, t)=0$ and $\fatv(\fatx, t)=\fatzero$.
Denote the unit vector with 1 in the $i$:th component by $\fate^i$.
At $t=t_\ast$ we have
\[
\tfatsigma(\fatv, q)\fatn=-\fate^1\delta(\fatx-\fatx_\ast)\delta(t-t_\ast)-\psi\fath, 
\]
in the boundary condition in \eqref{eq:FSadj}. 
For $t<t_\ast$, $\tfatsigma(\fatv, q)\fatn=-\psi\fath$. 
Since $\psi$ is small for $t<t_\ast$ (see Sect.~\ref{sec:annumFS}), the dominant part of the solution is $\fatv(\fatx, t)=\fatv_0(\fatx)\delta(t-t_\ast)$ for some $\fatv_0$. 
To simplify the notation in the remainder of this paper, a variable with the subscript $\ast$ is evaluated at $(\fatx_\ast, t_\ast)$ or if it is time independent at $x_\ast$.

When the slip coefficient at the ice base is changed by $\dC$, then the change in $u_{1\ast}$ is by \eqref{eq:FSgrad} 
\begin{equation}\label{eq:seasonu}
\begin{array}{rll}
    \du_{1\ast}&=\dL=\displaystyle{\int_0^T\int_{\Gamma_b} f(\fatT\fatu)\fatT\fatu\cdot\fatT\fatv\,\dC\, \d\fatx\, \d t}\\
      &\displaystyle{\approx\int_{\Gamma_b} f(\fatT\fatu)\fatT\fatu\cdot\fatT\fatv_0\,\dC(\fatx, t_\ast)\, \d\fatx.}
\end{array}
\end{equation}
In this case, the perturbation $\du_{1\ast}$ mainly depends on $\dC$ at time $t_\ast$ and the contribution from previous $\dC(\fatx, t),\, t<t_\ast,$ is small.

Let the height $h_\ast$ be measured at $\Gamma_s$. 
Then
\[
       F(h)=h(x, t)\delta(\fatx-\fatx_\ast)\delta(t-t_\ast), \; F_h=\delta(\fatx-\fatx_\ast)\delta(t-t_\ast), \; F_\fatu=\fatzero.
\]
The solution of the adjoint equation \eqref{eq:FSadj} with $\tfatsigma(\fatv, q)\fatn=-\psi\fath$ at $\Gamma_s$ for $\fatv(\fatx, t)$ is non-zero since $\psi(\fatx, t)\ne 0$ for $t<t_\ast$.
In a seasonal variation, there is a time dependent perturbation $\dC(\fatx, t)=\dC_0(\fatx) \cos(2\pi t/\tau)$ added to a stationary time average $C(\fatx)$. 
The time constant $\tau$ could be for example 1~a (year). 
Assume that $f(\fatT\fatu)\fatT\fatu\cdot\fatT\fatv$ is approximately constant in time (e.g. if $\fatu$ varies slowly, then $\psi\approx {\rm const}$ 
and $\fatv\approx {\rm const}$ for $t<t_\ast$. Then the observation at the ice surface varies as
\begin{equation}
\begin{array}{rll}\label{eq:seasonh}
  \dh_\ast&\displaystyle{=\dL=\int_0^T\int_{\Gamma_b}f(\fatT\fatu)\fatT\fatu\cdot\fatT\fatv\,\dC(\fatx, t)  \, \d\fatx\, \d t}\\
               &\displaystyle{\approx\int_0^{t_\ast}\cos(2\pi t/\tau)\, \d t\int_{\Gamma_b} f(\fatT\fatu)\fatT\fatu\cdot\fatT\fatv\,\dC_0 \, \d\fatx}\\
               &\displaystyle{=\frac{\tau}{2\pi}\sin(2\pi t_\ast/\tau)\int_{\Gamma_b} f(\fatT\fatu)\fatT\fatu\cdot\fatT\fatv\,\dC_0 \, \d\fatx}.
\end{array}
\end{equation}
When the friction perturbation $\dC$ is large at $t_\ast=0, \tau/2, \tau\ldots,$ the effect on $h_\ast$ vanishes.
If the middle of the winter is at $n\tau,\; n=0,1,2,\ldots$, then the middle of the summer is at $(n+1/2)\tau$. 
The friction is at its maximum in the winter and at its minimum in the summer when the meltwater introduces lubrication. 
There is no change of $h_\ast$ in the middle of the summer, $\delta h_\ast=0$, but $C+\dC$ has its lowest value then. 
If $h_\ast$ is measured in the summer and compared to a mean value $h(\fatx)$, then $\dh(\fatx, t_\ast)=0$ and the wrong conclusion would be drawn that there is no change in $C$ if the phase shift between $\dC$ and $\dh$ in \eqref{eq:seasonh} is not accounted for.  

\begin{figure}[htbp]
  \begin{center}
    \includegraphics[width=0.95\linewidth]{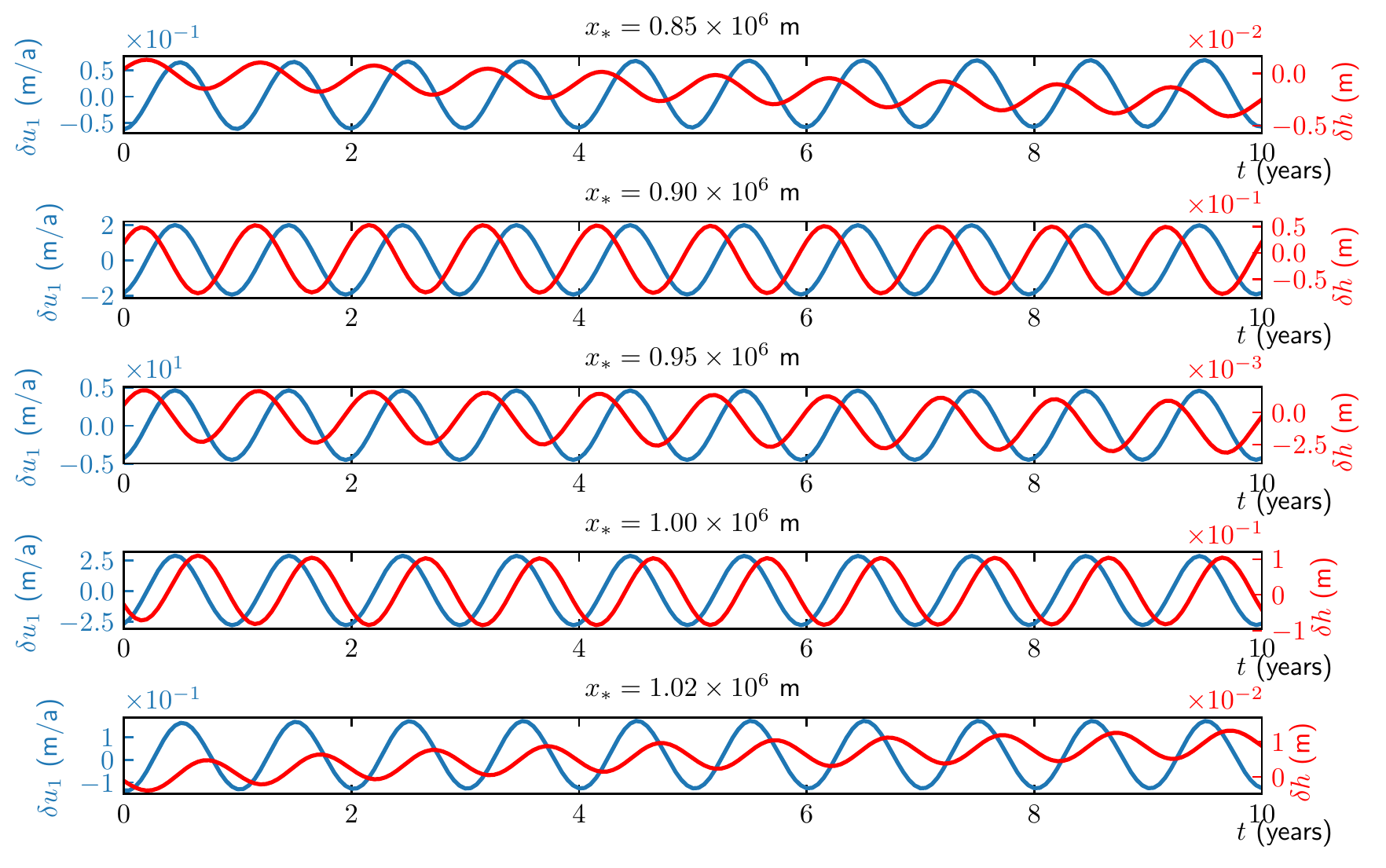}
  \end{center}
\caption{Observations at $x_\ast=0.85, 0.9, 0.95, 1.0, 1.02\times10^6$~m with FS in time $t\in[0,10]$ of $\du_1$(blue) and $\delta h$(red) with perturbation $\dC(t)=0.01C \cos(2\pi t)$ for $x\in[0.9, 1.0]\times10^6$~m. Notice the different scales on the $y$-axes.}
\label{fig:seasonalvariations}
\end{figure}

A two dimensional numerical example is shown in Fig.~\ref{fig:seasonalvariations} with $\tau=1$~a and $\delta C(x, t)=0.01C\cos(2\pi t)$ 
in an interval $x\in[0.9, 1.0]\times10^6$~m where the ice sheet flows from $x=0$ to $L=1.6\times10^6$~m. The grounding line is at $x_{GL}=1.035\times10^6$~m.
The details of the setup are found in the MISMIP \citep{MISMIP} test case used in \citep{CGPL19a}. The ice sheet is simulated by FS with Elmer/Ice \citep{ElmerDescrip} for 10 years. 
The perturbations in $\du_1$ and $\dh$ oscillate regularly with a period of 1 year after an initial transient and are small outside the interval $[0.9, 1.0]\times10^6$. 
An increase in the friction, $\dC>0$, leads to a decrease in the velocity and $\dC<0$ increases the velocity.
There is a phase shift $\Delta\phi$ in time by $\pi/2$ between $\du$ and $\dh$ as predicted by \eqref{eq:seasonu} and \eqref{eq:seasonh}.  
The weight in \eqref{eq:seasonh} for $\dC_0$ in the integral over $\fatx$ changes sign when the observation point is passing from $x_\ast=0.9\times 10^6$ to $1.0\times 10^6$ explaining why the shift changes sign in the two lower panels.
%The perturbations on $h$ behave as sine functions with positive coefficients on the left half of the perturbation area and negative coefficient on the right half.

The phase shift $\Delta\phi$ between the surface observations and the basal perturbations is investigated in \citep{Gudmundsson03} with a linearized equation and Fourier analysis. 
It is found that $\Delta\phi=-\pi/2$ between $\dC$ and $\dh$ for short perturbation wave lengths in the steady state as in Fig.~\ref{fig:seasonalvariations}.

%%%%%%%%%%%%%%%%%%%%%%%%%%%%%%%%%%%%%%%%%%%%%%%%%%%%%%%%%%%%%%%%%%%%%%%%%%%%%
%%%%%%%%%%%%%%%%%%%%%%%%%%%%%%%%%%%%%%%%%%%%%%%%%%%%%%%%%%%%%%%%%%%%%%%%%%%%%

%%%%%%%%%%%%%%%%%%%%%%%%%%%%%%%%%%%%%%%%%%%%%%%%%%%%%%%%%%%%%%%%%%%%%%%%%%%%%%%
%%%%%%% sensitivity and inverse FS
%%%%%%%%%%%%%%%%%%%%%%%%%%%%%%%%%%%%%%%%%%%%%%%%%%%%%%%%%%%%%%%%%%%%%%%%%%%%%
\subsubsection{The sensitivity problem and the inverse problem}\label{sec:invFSrel}

There is a relation between the sensitivity problem and the inverse problem to infer parameters from data.
Assume that $(\fatvInv^{i}, \qInv^i, \psiInv^i), i=1,\ldots, d,$ solves \eqref{eq:FSadj} with $F_\fatu=\fate^i\delta(\fatx-\fatxInv)$ or $F_h=\delta(\fatx-\fatxInv)$.
With $F_\fatu\ne 0, F_h=0$ we have $d=2\, (\text{or }3)$ in two (three)-dimensions and with $F_\fatu=0, F_h\ne 0$ we have $d=1$.
Consider a target functional $\calF$ for the steady state solution with weights $w_{i}(\fatxInv)$ multiplying $\du^i$ in the first variation of $\calF$.
Using \eqref{eq:FSgrad}, $\delta\calF$ is
\begin{equation}\label{eq:Fweights}
\begin{array}{rll}
    \delta\calF&=\displaystyle{\int_\omega \sum_{i=1}^d w_{i}(\fatxInv) \du^i\, \d \fatxInv
     =\int_\omega \sum_{i=1}^d w_{i}(\fatxInv) \int_{\Gamma_b}f(\fatT\fatu)\fatT\fatu\cdot\fatT\fatvInv^i\,\dC\,\d\fatx\, \d\fatxInv}\\
     &=\displaystyle{\int_{\Gamma_b}f(\fatT\fatu)\fatT\fatu\cdot\fatT\left(\int_\omega \sum_{i=1}^d w_{i}(\fatxInv)\fatvInv^i\, \d\fatxInv\right)\,\dC\,\d\fatx.}
\end{array}
\end{equation}
It follows from \eqref{eq:FSadj} that $(w_{i}(\fatxInv)\fatvInv^i(\fatx), w_{i}(\fatxInv)\qInv^i(\fatx), w_{i}(\fatxInv)\psiInv^i(\fatx))$ is a solution with
$F_\fatu=w_{i}(\fatxInv)\fate^i\delta(\fatx-\fatxInv)$ or $F_h=w(\fatxInv)\delta(\fatx-\fatxInv)$. Therefore, also 
\[
   \left(\int_\omega w_{i}(\fatxInv)\fatvInv^i\,\d\fatxInv, \int_\omega w_{i}(\fatxInv)\qInv^i\,\d\fatxInv, \int_\omega w_{i}(\fatxInv)\psiInv^i\,\d\fatxInv\right) 
\]
is a solution with 
$F_\fatu=\int_\omega w_{i}(\fatxInv)\fate^i\delta(\fatx-\fatxInv)\,\d\fatxInv=w_{i}(\fatx)\fate^i$ or $F_h=\int_\omega w(\fatxInv)\delta(\fatx-\fatxInv)\,\d\fatxInv=w(\fatx)$.

In the inverse problem, $\calF=\frac{1}{2}\int_\omega \|\fatu(\fatx)-\fatu_{\rm obs}(\fatx)\|^2\,\d\fatx$ \citep{Petra12} and the first variation is
$\delta\calF=\int_\omega (\fatu(\fatx)-\fatu_{\rm obs}(\fatx))\cdot\dfatu(\fatx)\,\d\fatx$. Let $w_i(\fatx)=u_i(\fatx)-u_{{\rm obs}, i}(\fatx)$ in \eqref{eq:Fweights}.
Then we find that
\begin{equation}\label{eq:dFtilde}
    \delta\calF=\int_{\Gamma_b} f(\fatT\fatu)\fatT\fatu\cdot\fatT\tilde{\fatv}(\fatx)\,\dC\,\d\fatx,
\end{equation}
where 
\[
     \tilde{\fatv}(\fatx)=\int_\omega \sum_{i=1}^d w_{i}(\fatxInv)\fatvInv^i(\fatx)\,\d\fatxInv
\]
is a solution to \eqref{eq:FSadj} with $F_\fatu=(w_1, w_2, w_3)^T=\fatu-\fatu_{\rm obs}$ or $F_h=w=h-h_{\rm obs}$. 

If we are interested in solving the inverse problem and determine $\delta\calF$ in \eqref{eq:Fweights} to iteratively compute the optimal solution with a gradient method, then we solve \eqref{eq:FSadj} directly with $F_\fatu=\fatu-\fatu_{\rm obs}$ or $F_h=h-h_{\rm obs}$ to obtain $\tilde{\fatv}$ without computing $\fatvInv^i$. 
  
%%%%%%%%%%%%%%%%%%%%%%%%%%%%%%%%%%%%%%%%%%%%%%%%%%%%%%%%%%%%%%%%%%%%%%%%%%%%%%%
%%%%%%% solutions FS
%%%%%%%%%%%%%%%%%%%%%%%%%%%%%%%%%%%%%%%%%%%%%%%%%%%%%%%%%%%%%%%%%%%%%%%%%%%%%
\subsubsection{Steady state solution to the adjoint height equation in two dimensions}\label{sec:annumFS}
% in Elmer/Ice according to \citep{Durand09b}, inverse methods in \citep{Fabien2012, ElmerDescrip}

In a two dimensional vertical ice, with $\fatu(x, z)=(u_1, u_3)^T$, the stationary equation for $\psi$ in \eqref{eq:FSadj} is 
\begin{equation}\label{eq:FS2Dstat}
    (u_1\psi)_x=F_h+(\fath\psi+F_\fatu)\cdot\fatu_z,\; z=h,\; 0\le x\le L.  
\end{equation}
When $x>x_\ast$, where $F_h=0$ and $F_\fatu=0$, then $\psi(x)=0$ since the right boundary condition is $\psi(L)=0$.

If $u_1$ is observed at $\Gamma_s$ then $F(\fatu, h)=u_1(x)\chi(x)$ and $F_\fatu=(\chi(x), 0)^T$ and $F_h=0$.
The weight $\chi$ on $u_1$ may be a Dirac delta, a Gaussian, or a constant in a limited interval.
On the other hand, if $F(\fatu, h)=h(x)\chi(x)$ then $F_h=\chi(x)$ and $F_\fatu=\fatzero$. 

Let $g(x)=u_{1z}(x)$ when $F_\fatu\ne \fatzero$ and let $g(x)=1$ when $F_h\ne 0$. 
Then by \eqref{eq:FS2Dstat}
\begin{equation}\label{eq:FS2Dstat2}
    (u_1\psi)_x-\fath\cdot\fatu_z\psi=g(x)\chi(x).  
\end{equation}
The solution to \eqref{eq:FS2Dstat2} is
\begin{equation}\label{eq:FSpsi}
\begin{array}{rll}
   \psi(x)&=\displaystyle{-\frac{1}{u_1(x)}\int_x^{x_\ast}\exp\left(-\int_x^{\xi}\frac{\fath\cdot\fatu_z(y)}{u_1(y)}\, \d y\right)g(\xi)\chi(\xi)\, \d\xi,\; 0\le x<x_\ast,}\\
   \psi(x)&=0,\; x_\ast<x\le L.
\end{array}
\end{equation}
In particular, if $\chi(x)=\delta(x-x_\ast)$ then $\calF=u(x_\ast)$ or $\calF=h(x_\ast)$ and the multiplier is
\begin{equation}\label{eq:FS2Dstat3}
   \psi(x)=-\frac{g(x_\ast)}{u_1(x)}\exp\left(-\int_x^{x_\ast}\frac{\fath\cdot\fatu_z(y)}{u_1(y)}\, \d y\right),\; 0\le x<x_\ast,
\end{equation}
which has a jump $-g(x_\ast)/u_1(x_\ast)$ at $x_\ast$.

With a small $\fath\cdot\fatu_z(y)\approx 0$ above, an approximate solution is $\psi(x)\approx -g(x_\ast)/u_1(x)$. Moreover, if $\fatu$ is observed with $F_\fatu\ne 0$ and $g(\fatx)\approx 0$, then $\psi(\fatx)\approx 0$ and $\psi\fath\approx 0$ in \eqref{eq:FSadj}. Consequently when $u_1$ is observed, the effect on $\fatv$ of the solution of the adjoint advection equation is negligible. It is sufficient to solve only the adjoint stress equation for $\fatv$ as in \citep{GCDGMMRR16, Isaac15, Petra12}, which may often be the case in FS. 
% For instance, we have in approximations of FS in SIA in \eqref{eq:SIAsol} $u_{1z}(x,y,h)=0$ and in SSA by assumption $u_{1z}(x)=0$ implying that $g(\fatx)\approx 0$. 

%%%%%%%%%%%%%%%%%%%%%%%%%%%%%%%%%%%%%%%%%%%%%%%%%%%%%%%%%%%%%%%%%%%%%%%%%%%%%%%
%%%%%%% Adjoint SSA
%%%%%%%%%%%%%%%%%%%%%%%%%%%%%%%%%%%%%%%%%%%%%%%%%%%%%%%%%%%%%%%%%%%%%%%%%%%%%
\subsection{Shallow shelf approximation} \label{sec:invSSA}

The adjoint equations for SSA are given and analyzed in this section.
A Lagrangian $\calL$ of the SSA equations is defined with the same technique as in \citep{Petra12} for the FS equations. 
By evaluating $\calL$ at the forward solution $(\fatu, h)$ and the adjoint solution $(\fatv, \psi)$, the effect of perturbed data at the ice base can be observed at the ice surface as a perturbation $\delta\calL$. 
The details of the derivations are found in Appendix~\ref{sec:appA}.2.
In a two dimensional vertical ice at the steady state, the equations are simpler and analytical solutions for the forward and adjoint equations are derived later in this section.

After insertion of the forward solution, partial integration in $\calL$, and applying the boundary conditions, the adjoint SSA equations are obtained as
\begin{equation}
\begin{array}{rll}\label{eq:SSAadj3D}
   &\psi_t+\fatu\cdot\nabla\psi+2\eta\fatB(\fatu):\fatD(\fatv)-\rho g H\nabla\cdot\fatv+\rho g \fatv\cdot\nabla b=F_h,\; \; {\rm in}\; \omega,\\
   &\psi(\fatx, T)=0,\; \;{\rm in}\;\omega,\quad \psi(\fatx, t)=0, \; \; {\rm on}\; \gamma_w,\\
   &\nabla\cdot\tfatvarsigma(\fatv)-Cf(\fatu)(\fatI+\fatF_\omega(\fatu))\fatv-H \nabla \psi=-F_\fatu,\quad  {\rm in}\;\omega,\\
   & \fatt\cdot\tfatvarsigma(\fatv)\fatn=-C_{\gamma}f_{\gamma}(\fatt\cdot\fatu)(1+F_\gamma(\fatt\cdot\fatu))\fatt\cdot\fatv,\;{\rm on}\; \gamma_g, 
     \quad \fatt\cdot\tfatvarsigma(\fatv)\fatn=0,\;{\rm on}\; \gamma_w,\\
   & \fatn\cdot\fatv=0,\;{\rm on}\; \gamma,
\end{array}
\end{equation}
where the adjoint viscosity $\tfateta$ and adjoint stress $\tfatvarsigma$ are defined by 
\begin{equation}
\begin{array}{rll}\label{eq:SSAviscadj3D}
   \tfateta(\fatu)&=\eta(\fatu)\left(\calI+\frac{1-n}{n \fatB(\fatu):\fatD(\fatu)}\fatB(\fatu)\otimes\fatD(\fatu)\right),\\ 
   \tfatvarsigma(\fatv)&=2H\tfateta(\fatu)\star\fatB(\fatv),
\end{array}
\end{equation}
cf. $\tfateta$ and $\tfatsigma$ of FS in \eqref{eq:FSviscadj}.
%The operator $\star$ is defined in \eqref{eq:stardef}. 
The adjoint equation derived in \citep{MacAyeal2} is the stress equation in \eqref{eq:SSAadj3D} with a constant $H$, $\fatF_\omega=0$ and $\tfateta(\fatu)=\eta(\fatu)$.

The adjoint SSA equations have the same structure as the adjoint FS equations \eqref{eq:FSadj}. 
There is one stress equation for the adjoint velocity $\fatv$ and one equation for the multiplier $\psi$ corresponding to the height equation in \eqref{eq:SSAforw3D}. 
However, the advection equation for $\psi$ in \eqref{eq:SSAadj3D} depends on the adjoint velocity $\fatv$ which leads to a fully coupled system for $\fatv$ and $\psi$.

The equations are solved backward in time with a final condition on $\psi$ at $t=T$. 
As in \eqref{eq:SSAforw3D}, there is no time derivative in the stress equation.
With a Weertman friction law, $f(\fatu)=\|\fatu\|^{m-1}$ and $f_\gamma(\fatt\cdot\fatu)=|\fatt\cdot\fatu|^{m-1}$, it is shown in Appendix~\ref{sec:adjvisc} that
\[
         \fatF_\omega(\fatu)=\frac{m-1}{\fatu\cdot\fatu}\fatu\otimes\fatu,\quad F_\gamma=m-1.
\]

If the friction coefficient $C$ at the ice base is changed by $\dC$, the bottom topography is changed by $\db$, and the lateral friction coefficient $C_\gamma$ is changed by $\dC_\gamma$, then it follows from Appendix~\ref{sec:appAdjointSSA} that the Lagrangian is changed by
\begin{equation}
\begin{array}{rll}\label{eq:SSAgrad3D}
  \delta\calL=&\displaystyle{\int_0^T\int_\omega (2\eta\fatB(\fatu):\fatD(\fatv)+\rho g\fatv\cdot\nabla h+\nabla\psi\cdot\fatu)\,\delta b
               -f(\fatu)\fatu \cdot\fatv\,  \delta C\,\,\d\fatx\, \d t}\\
             &\displaystyle{-\int_0^T\int_{\gamma_g} f_\gamma(\fatt\cdot\fatu)\fatt\cdot\fatu\,\fatt\cdot\fatv\,\dC_\gamma\, \d s\, \d t.}
\end{array}
\end{equation}
The weight in front of $\dC$ in \eqref{eq:SSAgrad3D} is actually the same as in \eqref{eq:FSgrad}.

Suppose that $h$ is observed with $F_\fatu=0$ in \eqref{eq:SSAadj3D}. Then the adjoint height equation must be solved for $\psi\ne 0$ to have a $\fatv\ne\fatzero$ in the
adjoint stress equation and a perturbation in the Lagrangian in \eqref{eq:SSAgrad3D}. The same conclusion followed from the adjoint FS equations.

The SSA model is obtained from the FS model in \citep{GreveBlatterBok, SSA} under some simplifying assumptions in the stress equation. 
An alternative derivation of the adjoint SSA would be to simplify the stress equation for $\fatv$ in the adjoint FS equations \eqref{eq:FSadj} under the same assumptions.
The resulting adjoint equation would be different from \eqref{eq:SSAadj3D} since the advection equation there depends on the adjoint velocity.

%%%%%%%%%%%%%%%%%%%%%%%%%%%%%%%%%%%%%%%%%%%%%%%%%%%%%%%%%%%%%%%%%%%%%%%%%%%%%
%%%%%%% Simplify to 2D problem
%%%%%%%%%%%%%%%%%%%%%%%%%%%%%%%%%%%%%%%%%%%%%%%%%%%%%%%%%%%%%%%%%%%%%%%%%%%%%
\subsubsection{SSA in two dimensions}\label{sec:annumSSA}

The forward and adjoint SSA equations in a two dimensional vertical ice are derived from \eqref{eq:SSAforw3D} and \eqref{eq:SSAadj3D} by letting $H$ and $u_1$ be independent of $y$ and taking $u_2=0$.  
Since there is no lateral force, $C_\gamma=0$. 
The position of the grounding line, where the ice starts floating on water, is denoted by $x_{GL}$ as in Fig.~\ref{fig:Icegeometry} and $\Gamma_b=[0, x_{GL}],\; \Gamma_w=(x_{GL}, L]$.
Let $C$ be a positive constant on the bedrock and $C=0$ on the water.
Simplify the notation with $u=u_1$ and $v=v_1$. 
The forward equation is
\begin{equation}
  \begin{array}{rll}\label{eq:SSAforw}
     &h_t+(uH)_x=a,\; \; 0\le t \le T,\; 0\le x\le L,\\
     &h(x, 0)=h_0(x),\; h(0,t)=h_L(t),\\
     &(H\eta u_x)_x-Cf(u)u-\rho g H h_x=0,\\
     &u(0,t)=u_l(t),\; u(L, t)=u_c(t),
  \end{array}
\end{equation}
where $u_l$ is the speed of the ice flux at $x=0$ and $u_c$ is the so-called calving rate at $x=L$. 
By the stress balance in \eqref{eq:SSAcf}, the calving front boundary satisfies 
\[
  u_x(L,t)=A\left[\frac{\rho g H(L, t)}{4}(1-\frac{\rho}{\rho_w})\right]^n.
\]

Assume that $u>0$ and $u_x>0$, then $\eta=2A^{-\frac{1}{n}} u^{\frac{1-n}{n}}_x$ and the friction term with a Weertman law is $Cf(u)u=Cu^m$. 
The adjoint equations for $v$ and $\psi$ follow either from simplifying the adjoint equations \eqref{eq:SSAadj3D} or deriving the adjoint equations from the two dimensional forward SSA \eqref{eq:SSAforw} 
\begin{equation}
\begin{array}{rll}\label{eq:SSAadj}
    &\psi_t+u\psi_x+(\eta u_x-\rho g H)v_x+\rho g b_x v=F_h,\quad 
    0\le t \le T,\; 0\le x\le L,\\
    &\psi(x, T)=0,\; \psi(L, t)=0,\\
   &(\frac{1}{n}\eta Hv_x)_x-C m f(u)v-H \psi_x=-F_u,\\
    &v(0, t)=0, \; v(L, t)=0.
\end{array}
\end{equation}
The coefficient $1/n$ in front of $\eta$ in the equation for $v$ is a result of the adjoint viscosity $\tfateta$,
which was $1$ in the adjoint SSA formulation in \citep{MacAyeal2}.

The effect on the Lagrangian of perturbations $\db$ and $\dC$ is obtained from \eqref{eq:SSAgrad3D} 
\begin{equation}
\label{eq:SSAgrad}
   \delta\calL=\int_0^T\int_0^L (\psi_x u+v_x\eta u_x+v \rho gh_x)\,\delta b-v f(u)u \,\delta C\,\,\d x\, \d t.
\end{equation}
The weights in the integral are denoted by $w_C=-v f(u)u$ and $w_b=\psi_x u+v_x\eta u_x+v \rho gh_x$.

%{\color{red}In the SSA model, the pertubation results are almost identical to the FS model when the friction coefficient is perturbed temporally as in Fig.~\ref{fig:seasonalvariations}.}

%%%%%%%%%%%%%%%%%%%%%%%%%%%%%%%%%%%%%%%%%%%%%%%%%%%%%%%%%%%%%%%%%%%%%%%%%%%%%
%%%%%%% Pertubation on b
%%%%%%%%%%%%%%%%%%%%%%%%%%%%%%%%%%%%%%%%%%%%%%%%%%%%%%%%%%%%%%%%%%%%%%%%%%%%%
\subsubsection{The forward steady state solution}\label{sec:SSAforwardSteady}

The viscosity terms in \eqref{eq:SSAforw} and \eqref{eq:SSAadj} are often small and the longitudinal stress can be ignored in the steady state solution, see \citep{Schoof07}.
The approximations of both the forward and adjoint equations can then be solved analytically on a reduced computational domain with $x\in[0,x_{GL}]$.
The simplified forward steady state equation in \eqref{eq:SSAforw} with $f(u)=u^{m-1}$ is written as
\begin{equation}\begin{array}{rll}\label{eq:SSAforwsimp}
   &(uH)_x=a,\;  0\le x\le x_{GL},\\
   &H(0)=H_0,\\
   &-Cu^m-\rho g H h_x=0,\\
   &u(0)=0,
\end{array}
\end{equation}
and the adjoint equation in \eqref{eq:SSAadj} is simplified when the gradient of the base topography $b_x$ is small 
\begin{equation}
\begin{array}{rll}\label{eq:SSAadjsimp}
   &u\psi_x-\rho g H v_x=F_h,\; \; 0\le x\le x_{GL},\\
   &\psi_x(0)=0,\;\psi(x_{GL})=0,\\
   &-Cmu^{m-1} v-H \psi_x=-F_u ,\\
   &v(0)=0.
\end{array}
\end{equation}
Numerical experiments in \citep{CGPL19a} show that an accurate solution compared to the FS and SSA solutions is obtained by calibration of $H$ with $H_{GL}=H(x_{GL})$ in \eqref{eq:SSAforwsimp}.
All the assumptions made for the simplification in \eqref{eq:SSAforwsimp} are not valid close to the ice divide at $x=0$ 
and the ice dynamics in this area cannot be captured accurately by SSA.

\begin{figure}[htbp]
  \begin{center}
    \includegraphics[width=0.95\linewidth]{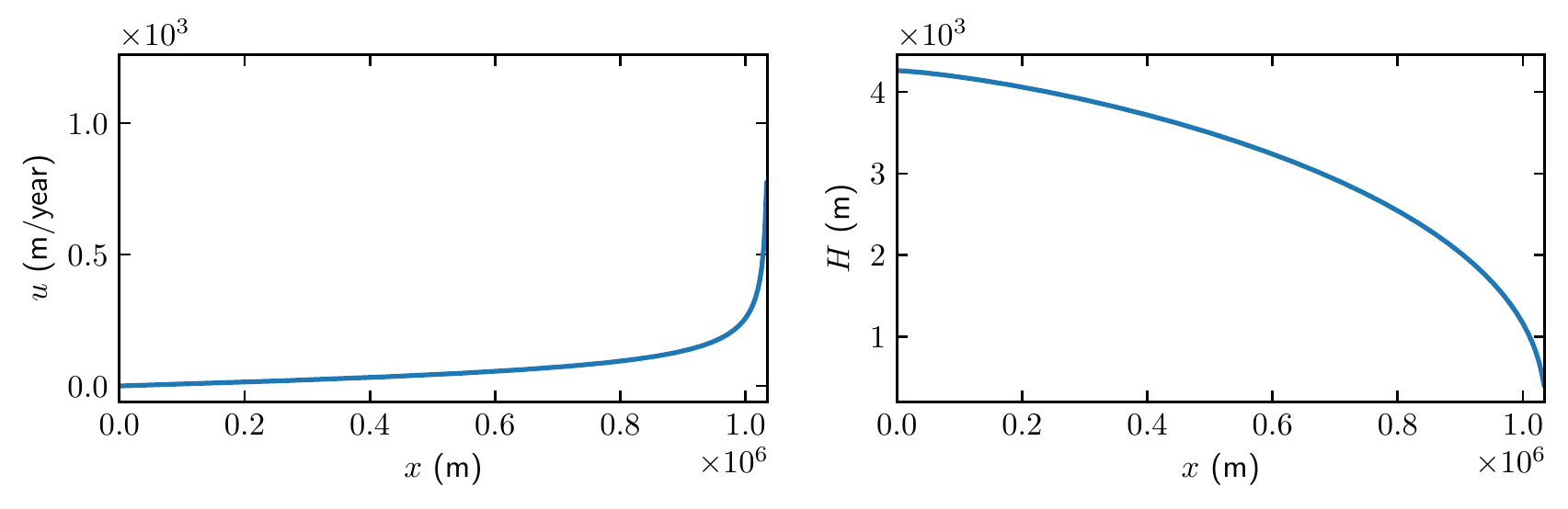}
  \end{center}
\caption{The analytical solutions $u(x)$ and $H(x)$ in \eqref{eq:SSA2Dsol} for a grounded ice in $[0, x_{GL}]$. }
\label{fig:uAndH}
\end{figure}

The solution to the forward equation \eqref{eq:SSAforwsimp} is determined when $a$ and $C$ are constant 
in \eqref{eq:SSAHsol} and \eqref{eq:SSAusol} in Appendix~\ref{sec:SSAanalyt} by integrating \eqref{eq:SSAintH} from $x$ to $x_{GL}$ 
\begin{equation}
\begin{array}{rll}\label{eq:SSA2Dsol}
   H(x)&=\displaystyle{\left(H^{m+2}_{GL}+\frac{m+2}{m+1}\frac{C a^m}{\rho g}(x^{m+1}_{GL}-x^{m+1} )\right)^{\frac{1}{m+2}}},\; 0\le x\le x_{GL},\\
   H(x)&=\displaystyle{H_{GL}, \; x_{GL}<x<L,}\\
   u(x)&=\displaystyle{\frac{ax}{H},\; 0\le x\le x_{GL}}, \quad u(x)=\displaystyle{\frac{ax}{H_{GL}},\;  x_{GL}<x<L}.
\end{array}
\end{equation}
An example of the analytical solutions $u(x)$ and $H(x)$ for $x\in[0,x_{GL}]$ is given in Fig.~\ref{fig:uAndH}, where the ice rests on a downward sloping bedrock with the grounding line position at $x_{GL}$ as in the MISMIP test cases in \citep{MISMIP}. 
The details and specified data of this example are found in \citep{CGPL19a}. When $x$ approaches $x_{GL}$, then $u$ increases and $H$ decreases rapidly in the figure. 
An alternative solution to \eqref{eq:SSAforw} when $x>x_{GL}$ is found in \citep{GreveBlatterBok} where the assumption is that $H(x)$ is linear in $x$.

%Analytically, there is no differences between integrating from $x=0$ and $x=x_{GL}$.
%However, numerical experiments show that integrating from $x=0$ lead to a wrong solution since 

%%%%%%%%%%%%%%%%%%%%%%%%%%%
% u and H
%%%%%%%%%%%%%%%%%%%%%%%%%%%

%%%%%%%%%%%%%%%%%%%%%%%%%%%%%%%%%%%%%%%%%%%%%%%%%%%%%%%%%%%%%%%%%%%%%%%%%%%%%
%%%%%%% Analytical solutions of the adjoint
%%%%%%%%%%%%%%%%%%%%%%%%%%%%%%%%%%%%%%%%%%%%%%%%%%%%%%%%%%%%%%%%%%%%%%%%%%%%%

\subsubsection{The adjoint steady state solution with $F_u\neq0$}

\begin{figure}[htbp]
  \begin{center}
    \includegraphics[width=0.95\linewidth]{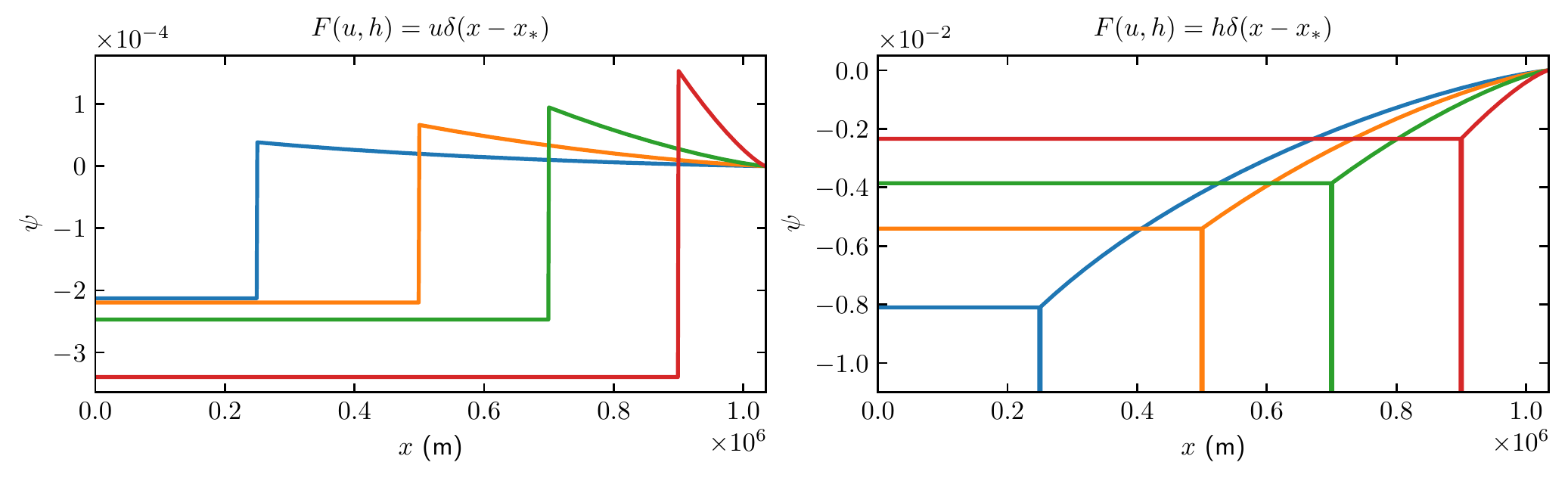}
  \end{center}
\caption{The analytical solutions of $\psi$ in \eqref{eq:SSAadjsimp} of the observations of $u$ (left panel) and $h$ (right panel) at different locations $x_\ast=0.25\times 10^6, 0.5\times 10^6, 0.7\times 10^6$ and $0.9\times10^6$~m (blue, orange, green and red).
% The expression of $\psi$ with respect to the observation of $u$ is in \eqref{eq:SSApsivsolU} and the $h$ response is in \eqref{eq:SSApsivsolH}. 
}
\label{fig:psi}
\end{figure}

\begin{figure}[htbp]
  \begin{center}
    \includegraphics[width=0.95\linewidth]{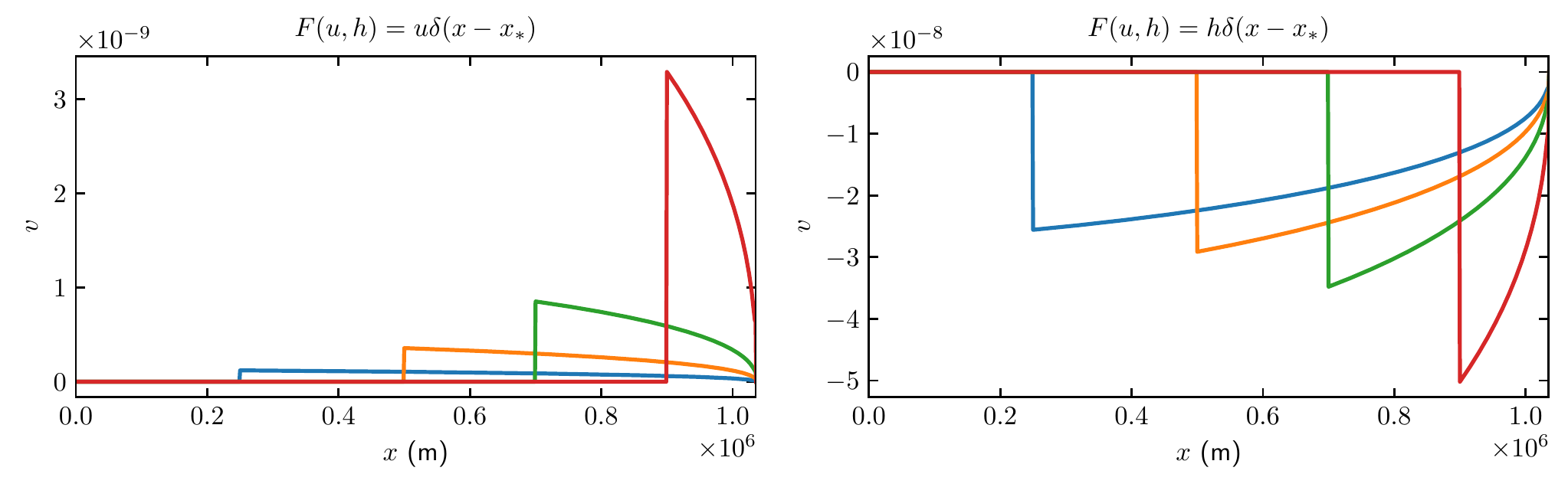}
  \end{center}
\caption{The analytical solutions of $v$ in \eqref{eq:SSAadjsimp} of the observations of $u$ (left panel) and $h$ (right panel) at different locations $x_\ast=0.25\times 10^6, 0.5\times 10^6, 0.7\times 10^6$ and $0.9\times10^6$~m (blue, orange, green and red).
% The expression of $v$ with respect to the observation of $u$ is in \eqref{eq:SSApsivsolU} and the continuous part of the $h$ response is in \eqref{eq:SSApsivsolH}. 
}
\label{fig:phi}
\end{figure}

The analytical solution to \eqref{eq:SSAadjsimp} is derived in Appendices~\ref{sec:SSAjumps} to \ref{sec:analyticalAdjointSSA} assuming that $b_x$ is small such that $b_x\ll H_x$ and $a$ and $C$ are constants.
The expressions for $\psi$ and $v$ are found in \eqref{eq:SSAvsol}-\eqref{eq:SSApsisol2}.
% {\color{red} A variable with the subscript $\ast$ is evaluated at $x_\ast$. }
When $u$ is observed at $x_\ast$, then $\calF$ and $F(u, h)$ satisfy 
\[
   \calF=\int_0^L u(x)\delta(x-x_{*})\, \d x=u_\ast,\;  F_u=\delta(x-x_{*}),\; F_h=0,
\]
and the adjoint solutions are
\begin{equation}
\begin{array}{rll}\label{eq:SSApsivsolU}
    \psi(x)&=\displaystyle{\frac{C a^m x_{*}}{\rho g H_{*}^{m+3}}\left(x_{GL}^m-x^m\right)},\; x_\ast< x \leq x_{GL},\\
     \psi(x)&=\displaystyle{-\frac{1}{H_\ast}+\frac{C a^m x_{*}}{\rho g H_{*}^{m+3}}\left(x_{GL}^m-x_\ast^m\right)},\; 0\le x< x_\ast,\\
       v(x)&=\displaystyle{\frac{a x_{*}}{\rho g H_{*}^{m+3}}H^m},\; x_\ast< x \le x_{GL},\\
       v(x)&=0,\; 0\le x< x_\ast,
\end{array}
\end{equation}
with discontinuities at the observation point $x_\ast$ in $\psi(x)$ and $v(x)$. 
The analytical solutions $\psi(x)$ and $v(x)$ of the ice sheet in Fig.~\ref{fig:uAndH} at different $x_\ast$ positions are shown in the left panels of Fig.~\ref{fig:psi} and Fig.~\ref{fig:phi}. 
In all figures, $m=1$ in the friction model. In Fig.~\ref{fig:psi} and subsequent figures, a Dirac term $\alpha \delta$ is plotted on a grid with grid size $\Delta x$ as a triangle with base $2\Delta x$ and height $\alpha/\Delta x$ such that the integral over the triangle is $\alpha$. The derivative $\alpha\delta'$ is plotted as a peak $\alpha/\Delta x^2$ followed by $-\alpha/\Delta x^2$. As in Sect.~\ref{sec:annumFS} for the FS model, $\psi$ can be small in SSA when $\fatu$ is observed compared to observations of $h$ (cf. left and right panels of Fig.~\ref{fig:psi}).

%%%%%%%%%%%%%%%%%%%%%%%%%%%
% psi
%%%%%%%%%%%%%%%%%%%%%%%%%%%
%%%%%%%%%%%%%%%%%%%%%%%%%%%
% v
%%%%%%%%%%%%%%%%%%%%%%%%%%%

\begin{figure}[htbp]
  \begin{center}
    \includegraphics[width=0.95\linewidth]{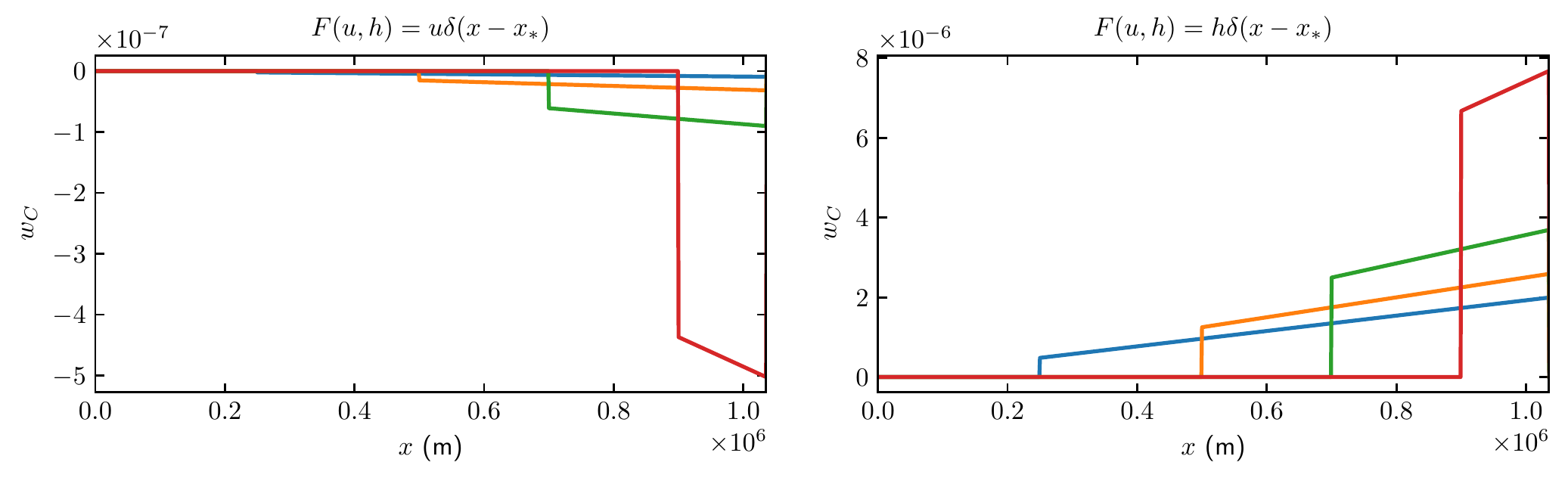}
  \end{center}
\caption{The analytical solution of the weights $w_C=-vu^m$ on $\dC$ in \eqref{eq:SSAgrad} for  $u$ (left panel) and $h$ (right panel) observed at $x_\ast=0.25\times 10^6, 0.5\times 10^6, 0.7\times 10^6$ and $0.9\times10^6$~m (blue, orange, green and red).
% The expressions are given in \eqref{eq:SSAvu}. 
}
\label{fig:dC}
\end{figure}

\begin{figure}[htbp]
  \begin{center}
    \includegraphics[width=0.95\linewidth]{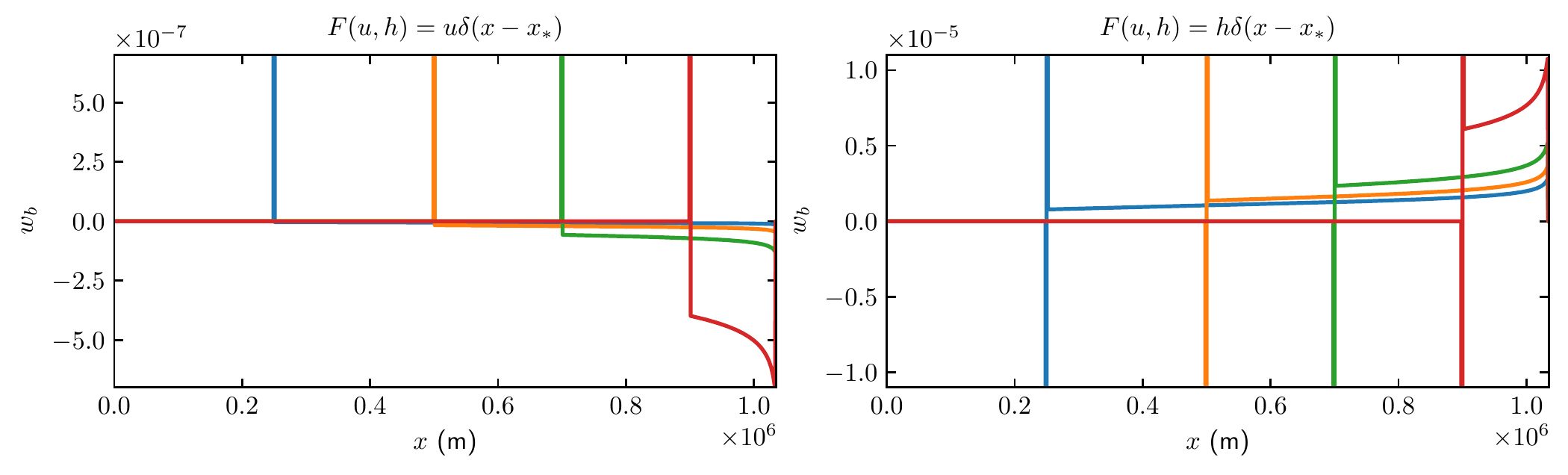}
  \end{center}
\caption{The analytical solution of weights $w_b=\psi_x u+v_x\eta u_x+v \rho gh_x$ on $\db$ in \eqref{eq:SSAgrad} for $u$ (left panel) and $h$ (right panel) observed at $x_\ast=0.25\times 10^6, 0.5\times 10^6, 0.7\times 10^6$ and $0.9\times10^6$~m (blue, orange, green and red).
% The expressions are given in \eqref{eq:SSAbweight}. 
}
\label{fig:db}
\end{figure}

The perturbation of the Lagrangian in \eqref{eq:SSAgrad} is derived in \eqref{eq:SSAvu}-\eqref{eq:SSAbweight} in Appendix~\ref{sec:analyticalAdjointSSA}
\begin{equation}\label{eq:SSAsimpL}
\begin{array}{lll}
  \displaystyle{\du_\ast}&=\delta\calL =\displaystyle{\int_0^{x_{GL}} (\psi_x u+v_x\eta u_x+v \rho gh_x)\,\delta b-v u^{m} \,\delta C\,\,\d x}\\
  &=\displaystyle{\int_{x_\ast^-}^{x_{GL}} \frac{ax_\ast{H^m}}{ H^{m+3}_{*}}\left[(m+1)H_x\mathcal{H}(x-x_\ast) +H\delta(x-x_\ast)\right] \,\delta b -\frac{ax_\ast(ax)^{m}}{\rho g H_{*}^{m+3}}\,\delta C\,\,\d x}\\
  &=\displaystyle{\frac{u_\ast\db_\ast}{H_\ast}-\frac{u_\ast}{\rho g H_\ast^{m+2}}\int_{x_\ast}^{x_{GL}} C (ax)^{m} \left((m+1)\frac{\db}{H} +\frac{\dC}{C}\right)\,\,\d x},
\end{array}
\end{equation}
or scaled with $u_\ast$
\begin{equation}\label{eq:SSAsimpL2}
\begin{array}{lll}
  \displaystyle{\frac{\du_\ast}{u_\ast}}=
  \displaystyle{\frac{\db_\ast}{H_\ast}-\frac{1}{\rho g H_\ast^{m+2}}\int_{x_\ast}^{x_{GL}} C (ax)^{m} \left((m+1)\frac{\db}{H} +\frac{\dC}{C}\right)\,\,\d x}.
\end{array}
\end{equation}
The Heaviside step function is denoted by $\mathcal{H}(x)$ in \eqref{eq:SSAsimpL}.
The weights with respect to $\dC$ and $\db$ in \eqref{eq:SSAsimpL} are shown in the left panels of Fig.~\ref{fig:dC} and Fig.~\ref{fig:db} for the same ice sheet geometry as in  Fig.~\ref{fig:uAndH}. 

%%%%%%%%%%%%%%%%%%%%%%%%%%%
% C weights
%%%%%%%%%%%%%%%%%%%%%%%%%%%

%%%%%%%%%%%%%%%%%%%%%%%%%%%
% b weights
%%%%%%%%%%%%%%%%%%%%%%%%%%%

The following conclusions can be drawn from \eqref{eq:SSAsimpL} and \eqref{eq:SSAsimpL2}:
\begin{enumerate}
\item The weight in front of $\dC$ increases when $x_\ast\rightarrow x_{GL}$. This is an effect of the increasing velocity $u_\ast$ and the decreasing thickness $H_\ast$, see Fig.~\ref{fig:uAndH}. 
Therefore, a perturbation $\dC$ with support in $[x_\ast, x_{GL}]$ will cause larger perturbations at the surface the closer $x_\ast$ is to $x_{GL}$ and the closer $\dC(x)$ is to $x_{GL}$.

\item If $\db_\ast=0$, i.e. the topography is unperturbed at the base below the observation point,  then the contribution of $\db$ in $(x_\ast, x_{GL}]$ cannot be separated from the contribution of $\dC$ in \eqref{eq:SSAsimpL2}.

\item The change in $u_\ast$ due to $\db$ in \eqref{eq:SSAsimpL} is simplified if $H_x\ll H$. Then $\du_\ast\approx u_\ast\db_\ast/H_\ast$ and the main effect on $u_\ast$ from the perturbation $\db$ is localized at each $x_\ast$.

\item The perturbation in $u$ is very sensitive to $H_\ast$ due to the factor $u_\ast/H_\ast^{m+2}\propto 1/H_\ast^{m+3}$ in \eqref{eq:SSAsimpL} when $x_\ast$ is moving downstream.

\item The relation between the two terms in \eqref{eq:SSAsimpL} is estimated in a case when $\dC=0$ and $\db/H$ is constant in $[x_\ast, x_{GL}]$. Let $x_\ast=0.9\cdot 10^6$ and $x_{GL}=1.2\cdot 10^6$. Then $u_\ast\approx 130$ in Fig.~\ref{fig:uAndH}. The factor multiplying $\db/H$ in the integral is by the left panel in Fig.~\ref{fig:db} approximately ${\rm weight}\times{\rm interval}\; {\rm length}\times H\approx 150$. The contribution by the two terms is about the same but with opposite signs. The integral term is reduced if the interval where $\dC\ne 0$ is shorter.
%The weight functions of $\db$ are shown in the same range as the weights of $\dC$ in Fig.~\ref{fig:dC}, although the coefficient multiplying the Dirac delta term in \eqref{eq:SSAbweight} is several magnitudes larger ($\sim 10^9$ in this case) which exceeds the ordinate limit of the figure.

\item With an unperturbed topography, let the perturbation of the friction coefficient be a constant $\dC_0\ne 0$ in $[x_0, x_1]\in[x_\ast, x_{GL}]$ resulting in a perturbation of the velocity. Evaluate the integral in \eqref{eq:SSAsimpL} to obtain
\begin{equation}\label{eq:SSAsimpL3}
  \displaystyle{\du_\ast}=\displaystyle{-\frac{u_\ast}{\rho g H_\ast^{m+2}}\int_{x_0}^{x_1} (ax)^{m} \dC_0\,\,\d x=-\frac{a^m u_\ast}{(m+1)\rho g H_\ast^{m+2}}(x_1^{m+1}-x_0^{m+1})\dC_0}.
\end{equation}
The same $\du_\ast$ is observed with a constant perturbation in $[x_2, x_3]\in[x_\ast, x_{GL}]$ with the amplitude $\dC_0(x_1^{m+1}-x_0^{m+1})/(x_3^{m+1}-x_2^{m+1})$. Different $\dC$ can give rise to the same observation $\du_\ast$. This holds true also for different $\db$ when $\db_\ast=0$. Inference of $\dC$ or $\db$ from surface data requires more observation points than one. 
\item Perturb $C$ by $\dC=\epsilon\cos(k x/x_{GL})$ in \eqref{eq:SSAsimpL} for some wave number $k$ and amplitude $\epsilon$ and let $\db=0$ and $m=1$. Then
\begin{equation}\label{eq:SSAsimpLk}
\begin{array}{lll}
  \du_\ast&=\displaystyle{-\int_{x_\ast}^{x_{GL}} \epsilon\frac{a^2 x_{*}}{\rho g H_{*}^{4}}x\cos\left(\frac{k x}{x_{GL}}\right)\,\,\d x}\\
         &=\displaystyle{-\epsilon\frac{a^2 x_{*}}{\rho g H_{*}^{4}}\frac{x_{GL}^2}{k}\left(\sin(k)-\frac{x_\ast}{x_{GL}}\sin\left(\frac{k x_\ast}{x_{GL}}\right)+\frac{1}{k}\left(\cos(k)-\cos\left(\frac{k x_\ast}{x_{GL}}\right)\right)\right)}.
\end{array}
\end{equation}
When $k$ grows at the ice base, the amplitude of the perturbation at the ice surface decays as $1/k$. 
The effect of high wave number perturbations of $C$ will be difficult to observe at the top of the ice.
When $k$ vanishes, then $\du_\ast$ tends to constant.
% This is in contrast to the SIA model in \eqref{eq:SIApert1} where all perturbations at the base, regardless of the wave number,  will be propagated undamped to the surface (see Section~\ref{sec:invSIA}).

\item If $\dC=0$ and $b$ is perturbed by $\db=\epsilon\cos(k x/x_{GL})$, then any perturbation at $x_\ast$ is propagated to the surface by the first term on the right hand side of  \eqref{eq:SSAsimpL}. The integral term will behave in the same way as in \eqref{eq:SSAsimpLk}. 
\end{enumerate}

Let the friction coefficient $C$ be perturbed by a constant $\dC$ in $[0, x_{GL}]$ and take $\db=0$. 
Then it follows from \eqref{eq:SSAsimpL} that
\begin{equation}\label{eq:SSAdu}
     \delta u_{*} =\dC \int_{x_\ast}^{x_{GL}} -v u^m\, \d x=\frac{a^{m+1}x_\ast(x_{*}^{m+1}-x_{GL}^{m+1})}{(m+1)\rho g H_{*}^{m+3}}\;\dC.%=\frac{u_{*}^{m+1}x_{*}}{(m+1)\rho g H_{*}^{2}}\;\dC.
\end{equation}
Computing the derivative of $u(x, C)$ with respect to $C$ in the explicit expression in \eqref{eq:SSA2Dsol} at $x_{*}$ yields the same result. 
%The impact of a constant perturbation $\dC$ at the ice base is growing rapidly, the closer to the grounding line the observation point is since $x_\ast$ is growing and $H_\ast$ is decreasing in \eqref{eq:SSAdu} when $x_\ast$ is approaching $x_{GL}$.

The sensitivity of surface data to changes in $b$ and $C$ is estimated in \citep{Gudmundsson08b} with a linearized model and Fourier analysis. 
The conclusion is that differences of short wavelength in the bedrock topography cannot be observed at the surface. Only differences with long wavelength in the friction coefficient propagate to the top of the ice. 
This is in agreement with \eqref{eq:SSAsimpL} and \eqref{eq:SSAsimpL2} and conclusions 7 and 8 above.

%%%%%%%%%%%%%%%%%%%%%%%%%%%%%%%%%%%%%%%%%%%%%%%%%%%%%%%%%%%%%%%%%%%%%%%%%%%%%
\subsubsection{The adjoint steady state solution with $F_h\neq0$}

In the case when $h$ is observed at $x_\ast$ and $F_u=0$ and $F_h=\delta(x-x_\ast)$, the expressions for $\psi$ and $v$ are 
\begin{equation}
\begin{array}{rll}\label{eq:SSApsivsolH}
    \psi(x)&=\displaystyle{-\frac{C a^{m-1} }{\rho g H_{*}^{m+1}}\left(x_{GL}^m-x^m\right)},\; x_\ast< x \leq x_{GL},\\
     \psi(x)&=\displaystyle{-\frac{C a^{m-1} }{\rho g H_{*}^{m+1}}\left(x_{GL}^m-x_\ast^m\right)},\; 0\le x< x_\ast,\\
       v(x)&=\displaystyle{-\frac{H^m}{\rho g H_{*}^{m+1}}},\; x_\ast< x \le x_{GL},\\
       v(x)&=0,\; 0\le x< x_\ast.
\end{array}
\end{equation}
There is a discontinuity at the observation point $x_\ast$ in  $v(x)$, see  Fig.~\ref{fig:phi}, but  $\psi(x)$ is continuous in the solution of \eqref{eq:SSAadjsimp}. 
Actually, $\psi(x)\sim-\delta(x-x_\ast)$ in the neighborhood of $x_\ast$ due to the second derivative term $(\frac{1}{n}\eta Hv_x)_x$ which is neglected in the simplified equation \eqref{eq:SSAadjsimp} but is of importance at $x_\ast$. 
A correction $\hat\psi$ of $\psi$ at $x_\ast$ is therefore introduced to satisfy $\left(\frac{1}{n}\eta H v_x\right)_x-H\hat\psi_x = 0$. 
With $v_x(x_\ast)=-\delta(x-x_\ast)/(\rho g H_\ast)$, the correction is $\hat\psi(x)=-\delta(x-x_\ast)\eta_\ast/(n\rho g H_\ast)$. 
The solution $\psi$ is corrected at each $x_\ast$ in Fig.~\ref{fig:psi} with $\hat\psi$.
The perturbation in $h$ is as in \eqref{eq:SSAsimpL} with $\psi$ and $v$ in \eqref{eq:SSApsivsolH} and the additional term $\hat\psi$
\begin{equation}\label{eq:SSAsimph}
  \begin{array}{lll}
    \displaystyle{\frac{\dh_\ast}{H_\ast}}&=\displaystyle{\int_{x_\ast^-}^{x_{GL}}-\frac{u\eta_\ast}{n \rho g H_\ast^2}\delta_x(x-x_\ast)\db}\,\,\d x+\displaystyle{\int_{x_\ast}^{x_{GL}} \frac{C (ax)^m}{\rho g H^{m+2}_{*}}\left((m+1)\frac{\db}{H} +\frac{\dC}{C}\right)\,\,\d x}\\
    &=\displaystyle{\frac{a\eta_\ast}{n\rho g H_\ast^2}\left(x\frac{\db}{H}\right)_x(x_\ast)+{\frac{1}{\rho g H^{m+2}_{*}}\int_{x_\ast}^{x_{GL}} C (ax)^m\left((m+1)\frac{\db}{H} +\frac{\dC}{C}\right)\,\,\d x}},
  \end{array}
\end{equation}
where $a(x\db/H)_x(x_\ast)=(u\db)_x(x_\ast)$ represents the $x$-derivative of $u\db$ evaluated at $x_\ast$. When $\db=0$ then $\du_\ast$ in \eqref{eq:SSAsimpL2} 
and $\dh_\ast=\dH_\ast$ in \eqref{eq:SSAsimph} satisfy $\du_\ast H_\ast=-\dH_\ast u_\ast$ as in the integrated form of the advection equation 
in \eqref{eq:SSAforwsimp} and in \eqref{eq:SSAuHsol}. 

The contribution from the integrals in \eqref{eq:SSAsimpL2} and \eqref{eq:SSAsimph} is identical except for the sign (see also Fig.~\ref{fig:dC}). The first term in \eqref{eq:SSAsimpL2} depends on $\db/H$ and the first term in \eqref{eq:SSAsimph} depends on the derivative of $x\db/H$. Because of these similarities, the conclusions 1, 2, 6, 7, and 8 from \eqref{eq:SSAsimpL} and \eqref{eq:SSAsimpL2} for $\du_\ast$ are valid also for $\dh_\ast$ in \eqref{eq:SSAsimph}. The change in $h$ caused by $\dC$ is less sensitive to $H_\ast$ than the change in $u$ since the factor multiplying the integral is proportional to $1/H_\ast^{m+1}$.

%The adjoint variable $v$ and the weights for $\dC$ in the right panels where $F_h\ne 0$ of Fig.~\ref{fig:phi} and Fig.~\ref{fig:dC} behave in the same manner as in the left panels where $F_u\ne 0$ but with a change of sign.

%In Fig.~\ref{fig:db}, as $u\psi_x$ dominates the weights for $\db$ in \eqref{eq:SSAsimph},  $\hat\psi$ contributes with a term proportional to $-\delta_x(x-x_\ast)$ at $x_\ast$ whose coefficient is several magnitudes ($\sim 10^8$ in this case){\bf ???} larger than the remaining weight function.
%Therefore, the main perturbation in $h$ is due to the $x$-derivative of $u\db$ at $x_\ast$.{\bf ???}
If $u\db=ax\db/H$ is constant, then $\dh$ is determined by the continuous weight in $x\in[x_\ast, x_{GL}]$ in Fig.~\ref{fig:db} which has a shape similar to the $\dC$ weight in Fig.~\ref{fig:dC}. A perturbation of $b$ at the base is directly visible locally in $u$ at the surface while the effect of $\dC$ is non-local in \eqref{eq:SSAsimph}.

%%%%%%%%%%%%%%%%%%%%%%%%%%%%%%%%%%%%%%%%%%%%%%%%%%%%%
\subsubsection{The time dependent adjoint solution}
%%%%%%%%%%%%%%%%%%%%%%%%%%%%%%%%%%%%%%%%%%%%%%%%%%%%%

% Settings: forward, adjoint
The time dependent adjoint equation \eqref{eq:SSAadj} is solved numerically for the MISMIP test case in Sect. \ref{sec:SSAforwardSteady}.
Starting from the steady state solution, the friction coefficient $C$ has a seasonal variation in the forward equation \eqref{eq:SSAforw}, such that $C(x, t)=C_0(1+\kappa\cos(2\pi t)),\; 0<\kappa<1$.
Apparently, $C$ has its highest value at $t=n,\; n=0, 1, 2,\ldots$, i.e. the winter, and its lowest value at $t=n+1/2$, i.e. the summer.
The period of the seasonal variation is 1~a and the beginning of each year is in the winter.

\begin{figure}[htbp]
  \begin{center}
    \includegraphics[width=0.95\linewidth]{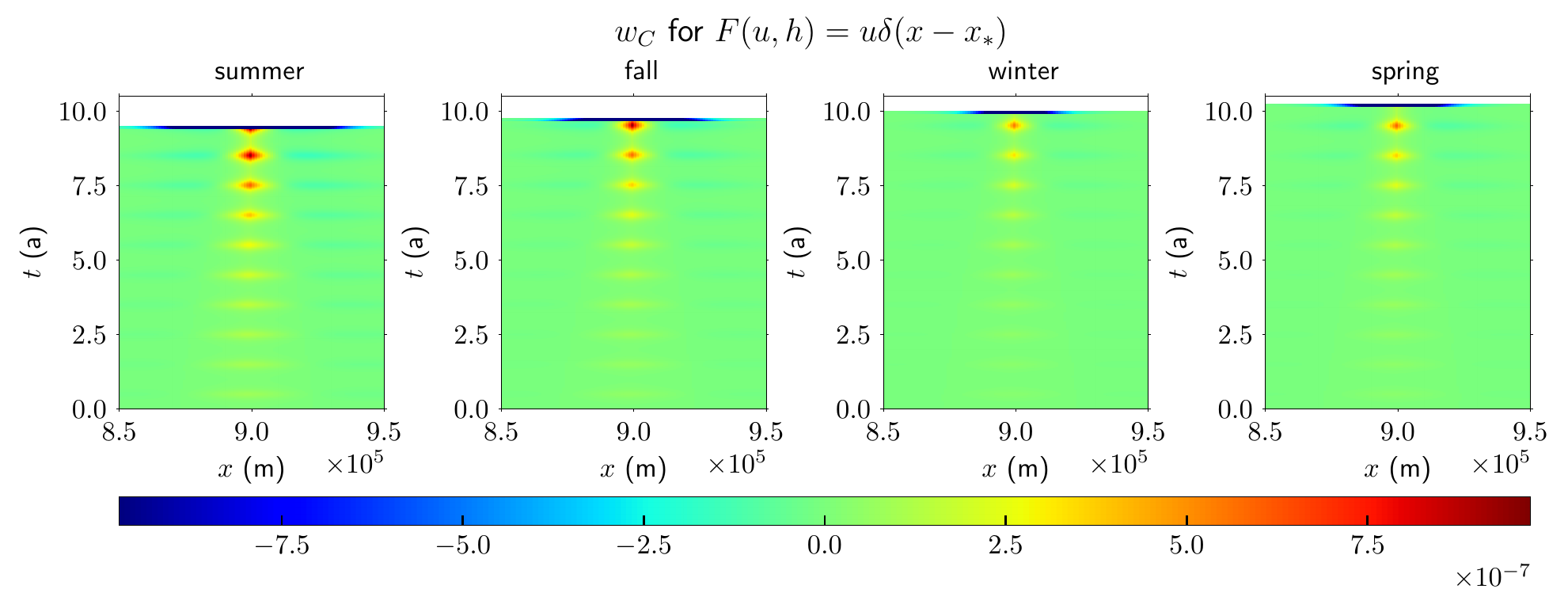}
    \includegraphics[width=0.95\linewidth]{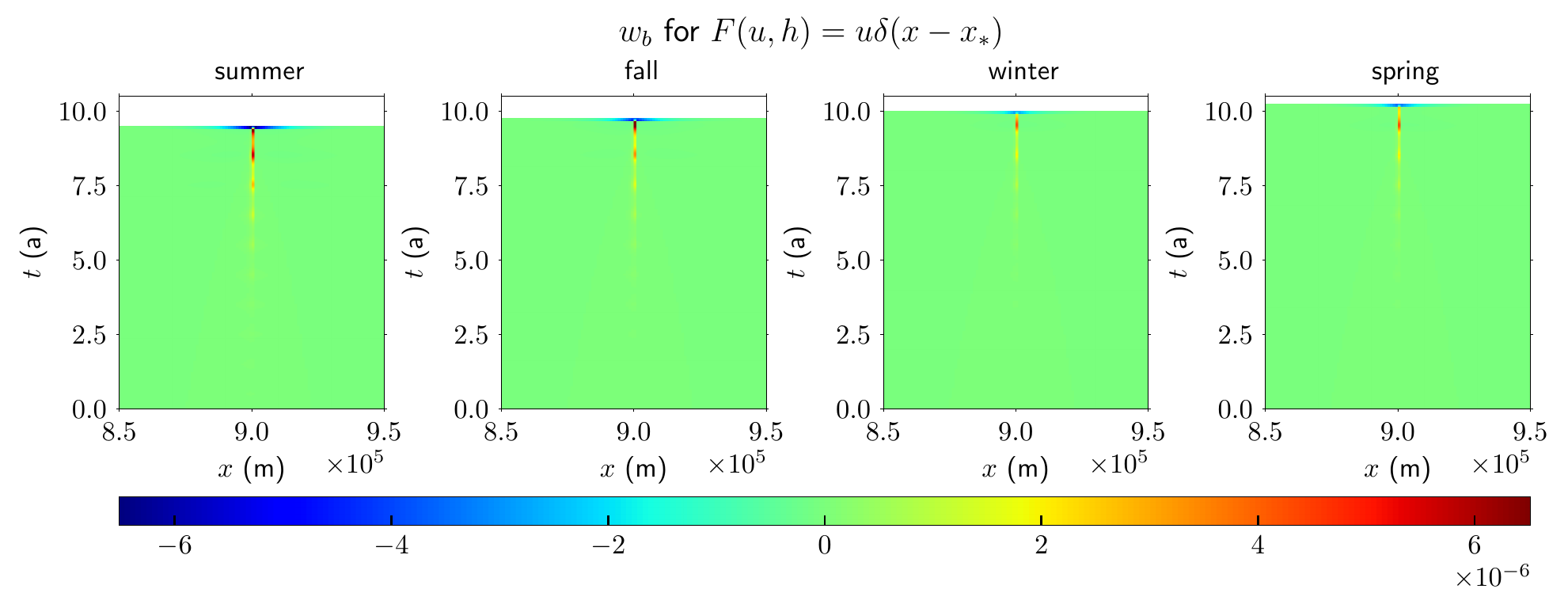}
    \includegraphics[width=0.95\linewidth]{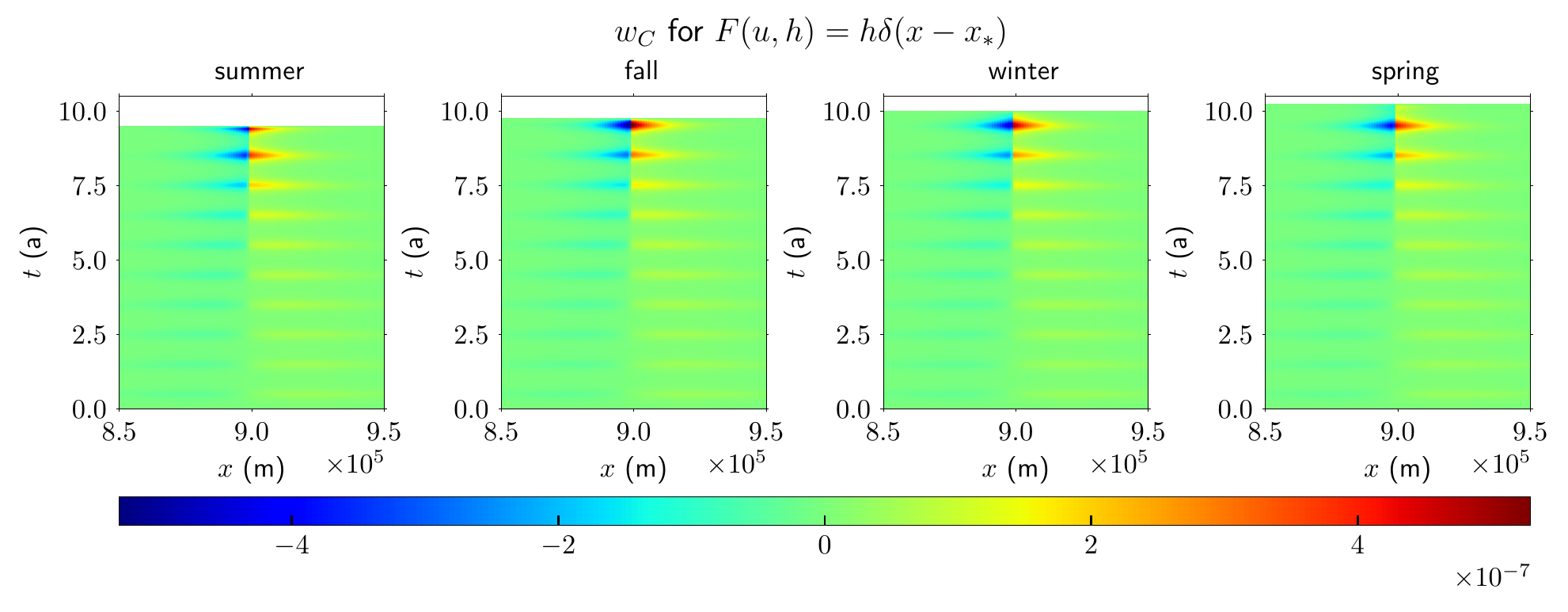}
    \includegraphics[width=0.95\linewidth]{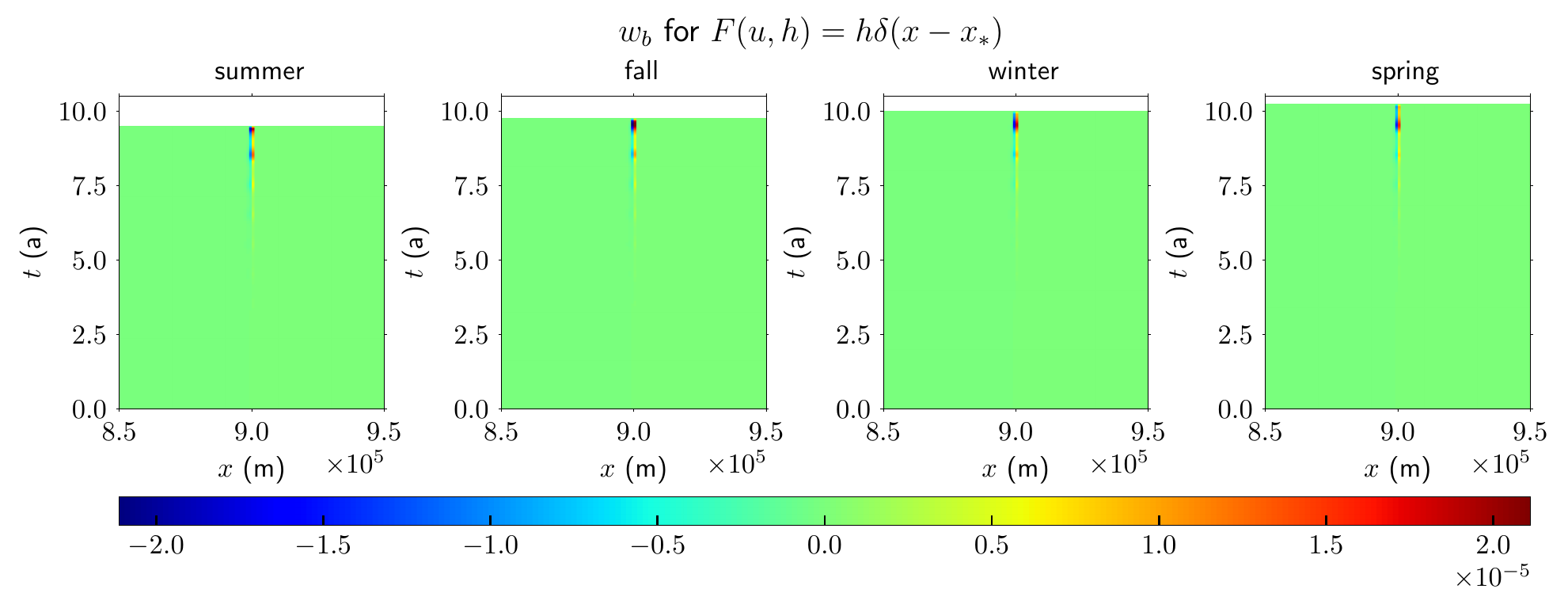}
  \end{center}
\caption{The adjoint weights for the observations at $x_*=9\times10^5$~m of the four seasons. The first row: $w_C$ for the observation of $u$.  The second row: $w_b$ for the observation of $u$.  The third row: $w_C$ for the observation of $h$.  The last row: $w_b$ for the observation of $h$.}
\label{fig:SSAAdjointTimeSpace}
\end{figure}

\begin{figure}[htbp]
  \begin{center}
    \includegraphics[width=0.95\linewidth]{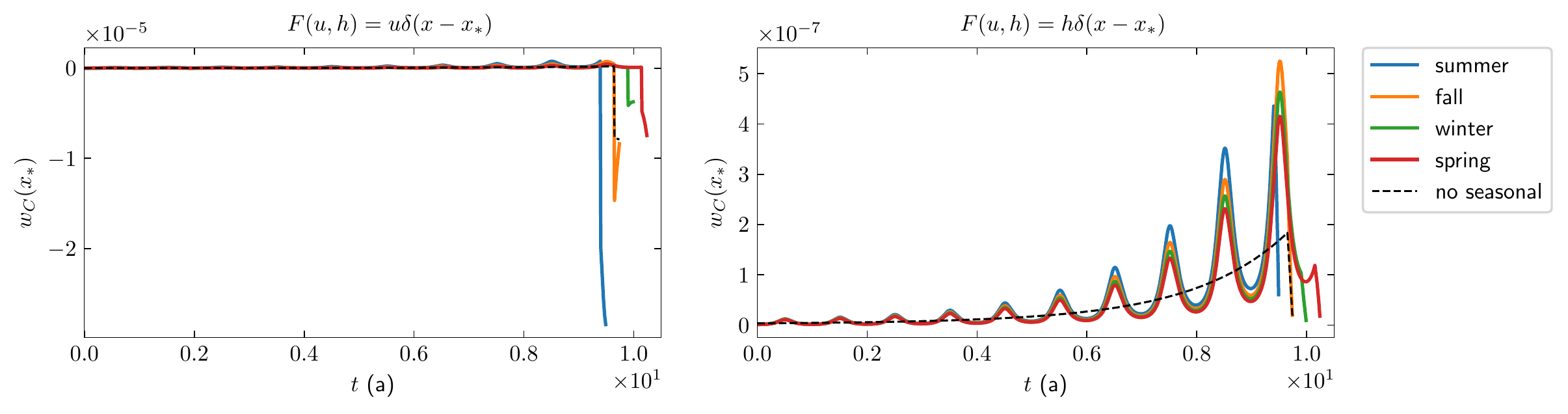}
    \includegraphics[width=0.95\linewidth]{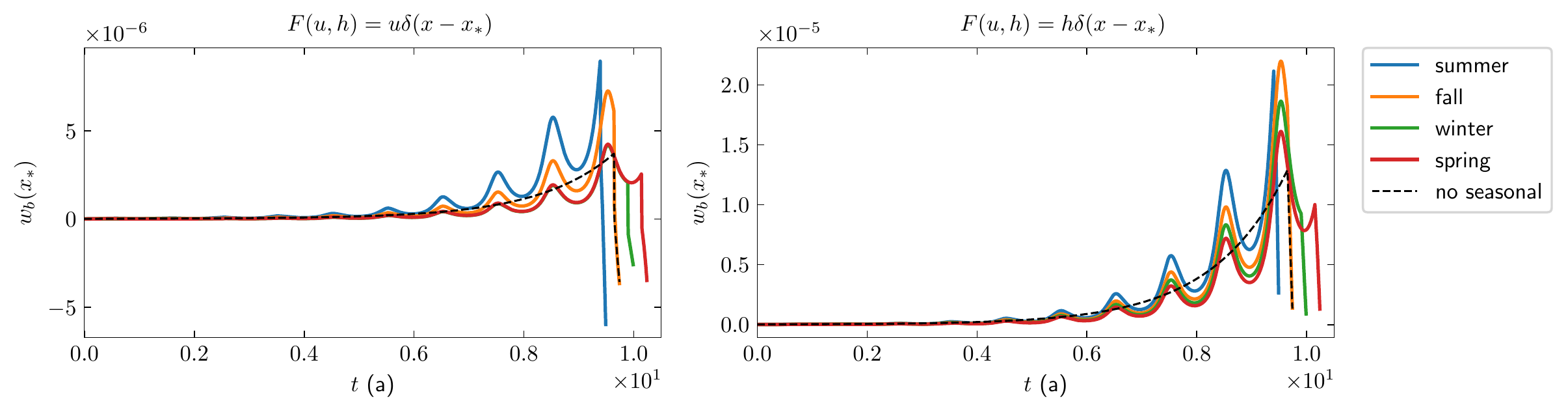}
  \end{center}
\caption{The adjoint weights for the observations of $u$ and $h$ at $x_\ast$ in the four seasons of the tenth year with seasonal varying friction coefficient. The black dashed line is a reference solution without seasonal variations which is observed at $t=9.75$.}
\label{fig:SSAAdjointTime}
\end{figure}

% numerical settings: dt=0.01a, dx=1km
The amplitude of the pertubation is set to $\kappa=0.5$ and the forward equation \eqref{eq:SSAforw} is solved for 11 years.
Observations on $u$ and $h$ are taken at $x_*=9\times10^5$~m for 0.1~a in the four seasons of the tenth year, e.g., in the summer $(t=9.5)$, the fall $(t_\ast=9.75)$, the winter $(t=10)$, and the spring $(t=10.25)$.
The adjoint equations \eqref{eq:SSAadj} are solved from the observation points backward in time, respectively, as in Figs.~\ref{fig:SSAAdjointTimeSpace} and \ref{fig:SSAAdjointTime}.
% In order to resolve the variation in time, 
According to a convergence test, the time step is chosen to be $0.01$~a and the spatial resolution is $10^3~$m.

%%%%% results

% In space-time
The adjoint weights $w_C$ and $w_b$, as defined in \eqref{eq:SSAgrad}, are shown in Fig.~\ref{fig:SSAAdjointTimeSpace} for the observations on $u$ and $h$ at $x_*=9\times10^5$~m in the four seasons.
The time axis in the figure starts from the steady state solution and follows the time direction in the forward problem.
As expected, most of the weights in space and time are negligible.
Therefore, we take a snapshot with the width of $10^5$~m in space around $x_\ast$.
Only $\dC$ and $\db$ in a narrow interval around $x_\ast$ for $t$ in $[0, t_\ast]$ have an influence on $\du_\ast$ and $\dh_\ast$, see Fig.~\ref{fig:SSAAdjointTimeSpace}. 
A perturbation at the base is propagated to the $x_\ast$ position on the surface but with a possible delay in time.
% Only $\dC$ and $\db$ close to $x_\ast$ influence $\du_\ast$ and $\dh_\ast$ in all the cases.

%%%%%%%%%%%%%%%%%%%%%%%%%%%
%%%%%%%%%%%%%%%%%%%%%%%%%%%

% W in time
The temporal variations of the adjoint weights at $x_\ast$ in Fig.~\ref{fig:SSAAdjointTimeSpace} are shown in Fig.~\ref{fig:SSAAdjointTime} for the four seasons with four different colors.
As expected, the weights vanish when $t>t_\ast$.
% Only $\dC$ and $\db$ close to $x_\ast$ and $t_\ast$ influence $\du_\ast$. 
% w_Cu: the influence from $\dC$ and $\db$ to $\du_\ast$ are mostly concentrated at $t_\ast$.
In the left panels of Fig.~\ref{fig:SSAAdjointTime}, the perturbations $\dC_\ast$ and $\db_\ast$ have a direct effect on $\du_\ast$ at $t_\ast$, where $w_{C}(x_\ast)$ and $w_{b}(x_\ast)$ are both negative.
The same conclusion is valid for $\du_{1\ast}$ computed from the FS equations \eqref{eq:seasonu} in Sect. \ref{sec:timedepFS}.
A change in $\dC_\ast$ at the base is observed immediately as a change in $u$ at the surface. 
The effect of $\dC$ on $\du_\ast$ for $t<t_\ast$ is weak in the upper left panel of Fig.~\ref{fig:SSAAdjointTime}. 
The largest effect of $\dC$ on $\du_\ast$ and $\dh_\ast$ appears in the summer when $C$ is small, as the blue lines in the left panels.
% For the same $\du_\ast$ and $\dh_\ast$, $\dC$ and $\db$ have to be larger in the winter and smaller in the summer. 

% w_Ch and w_bh
However, when $h$ is observed, the effect of $\dC_\ast$ and $\db_\ast$ are not visible directly because $w_{C\ast}\approx 0$ and $w_{b\ast}\approx 0$ in the right panels of Fig.~\ref{fig:SSAAdjointTime}.
Additionally, the effect of $\dC$ and $\db$ is difficult to be separated, since the weight $w_{b}(x_\ast)$ has the similar shape as $w_{C}(x_\ast)$.
% phase shift and relations in different seasons
The largest effect on $\dh_\ast$ is from $\dC$ in the summer due to the peaks as in the upper right panel.
For the same $\dC$, the largest $\dh_\ast$ is observed in the fall (orange), then the second largest $\dh_\ast$ is in the winter (green) followed by the spring observation (red). 
If $\dh_\ast$ is observed in the fall and the time dependency is ignored, then the wrong conclusion is drawn that $\dC$ in the fall has the strongest effect. There is a delay in time between the perturbation and the observation of the effect.
%{\color{red} The phase shift in time $\Delta \phi=-\pi/2$ is similar to what we found with FS in \eqref{eq:seasonu} and Fig.~\ref{fig:seasonalvariations}. 
%}

% reference solution
A reference adjoint solution observed during the fall season ($t_\ast=9.75$) with $\kappa=0$ in the forward equations is shown in the black dashed lines in Fig.~\ref{fig:SSAAdjointTime}.
The weight $w_{b}$ for a constant $b$ in time is well approximated by $\exp(-(T-t)/\tau)$ with $\tau=1.4$~a. 
For the weight $w_C$, the same exponential function holds, but the constant $\tau=1.8$~a for the observation of $h_\ast$ and $\tau=2.2$~a for the $u_\ast$ case.
Suppose that the temporal perturbation is oscillatory $\dC_0\cos(2\pi f t)$ with frequency $f$. Then perturbation in $h$ at $t=t_\ast$ is
\begin{equation}\label{eq:adjsolappr}
 \begin{array}{rll}
 \dh&=\displaystyle{\int_0^T \exp(-(T-t)/\tau)\dC_0\cos(2\pi f t)\, \d t}\\
      &=\left(\displaystyle{\frac{\cos(2\pi f T)+2\pi\tau f\sin(2\pi f T)-e^{-T/\tau}}{4\pi^2\tau f^2+\tau^{-1}}}\right)\dC_0,
      % &=\left(\displaystyle{\frac{\sin(2\pi f T)}{\tau^{-1}+4\pi^2\tau f^2}
      % -\frac{\cos(2\pi f T)-\exp(-T/\tau)}{2\pi f+(2\pi f \tau^2)^{-1}}}\right)\dC_0,
\end{array}
\end{equation}
cf. \eqref{eq:SSAsimpLk}. With a high frequency (diurnal), $f\gg 1$, then $\dh\propto 1/f$ and high frequency perturbations are damped efficiently. If the frequency is low (decennial), $f\ll 1$, then $\dh\propto\tau$ and the change in $h$ is insensitive to the frequency. The same conclusions hold true for $\db$ where decennial variations seem more realistic.

%%%%%%%%%%%%%%%%%%%%%%%%%%%
%%%%%%%%%%%%%%%%%%%%%%%%%%%

\section{Conclusions}\label{sec:concl}
%%%%%%%%%%%%%%%%%%%%%%%%%%%%%%%%%%%%%%%%%%%%%%%%%%%%%%%%%%%%%%%%%%%%%%%%%%%%%%%%%%%%%%%%%%%%%%%%%%%%%%%%%%%%%%%%%%%%%%%%%%%%%%%%%%%%%%%%%%%%%%%%%%%%%
The sensitivity of the flow of ice to basal conditions is analyzed for time dependent and steady state solutions 
of the FS and SSA equations including the advection equation for the height.
Perturbations at the base of the ice are introduced in the friction coefficient $C$ and the topography $b$.
The analysis relies on the adjoint equations of the FS and SSA stress and advection equations. 
The adjoint FS and SSA equations follow from the Lagrangians of the forward FS and SSA equations.

The adjoint equations are derived for observations of the velocity $\fatu$ and height $h$ at the top surface of the ice.
The adjoint height equation in the FS model is solved analytically for a two dimensional vertical ice.
The relation between the inverse problem to find parameters from data and the sensitivity problem is established. The same adjoint equations are solved in the inverse problem but with other forcing functions. 

If the perturbations in the basal conditions are time dependent then time cannot be ignored in the inversion. It is necessary to include the adjoint height equation if $h$ is observed. The wrong conclusions may be drawn with only static snapshots for both the FS and the SSA model. If $\fatu$ is observed then the contribution of the solution of the adjoint height equation is small and it is sufficient to solve only the adjoint stress equation as is the case in many articles on inversion e.g. \citep{GCDGMMRR16,  Isaac15, Petra12}.

The adjoint equations of the FS and SSA models are similar and the analytical solutions based on the SSA equations for a two dimensional vertical ice show that the sensitivity grows as the 
observation point $x_\ast$ is approaching the grounding line separating the grounded and the floating parts of the ice. 
The reason is that the velocity increases and the thickness of the ice decreases.

In the steady state solution of SSA, there is a non-local effect of a perturbation $\dC$ in $C$ in the sense that $\dC(x)$ affects both $u(x_\ast)$ and $h(x_\ast)$ even if $x\ne x_\ast$, 
but $\db$ has a strong local effect concentrated at $x_\ast$.
Nevertheless, the shapes of the two sensitivity functions for $\db$ and $\dC$ are very similar except for the neighborhood of $x_\ast$.
It is possible to separate the effect of $\db$ and $\dC$ in the steady state SSA model thanks to the localized influence of 
$\db$ in $\du$ and $\dh$.
%in \eqref{eq:SSAsimpL} and \eqref{eq:SSAsimph}.
The same effect on $\du$ and $\dh$ at one observation point can be achieved by different $\dC$ and $\db$.
%In the SSA solution, an increase of the friction coefficient $C$ at $x_\ast$ leads to a decrease in $u$ but an increase in $h$ in the domain downstream of $x_\ast$. 
%On the other hand, an increase of the bedrock elevation $b$ at $x_\ast$ causes an increase in $u$ at the same position, but in the downstream direction, 
%it reduces  the velocity $u$. 
Perturbations in $C$ at the base observed in $u$ at the surface are damped inversely proportional to the wavenumber of $\dC$ 
thus making high wavenumber perturbations difficult to register at the top.

% {\color{red}{\bf remove ??}} The sensitivity in $u$ in the SIA model is local to basal perturbations in the sliding coefficient.

In the time dependent solution of SSA, the strongest impact of a perturbation at $x_\ast$ is made at $x_\ast$ at the surface possibly with a time delay. There is a local effect of $\dC$ and $\db$ at $(x_\ast, t_\ast)$ on $u_\ast$ but not on $h_\ast$ where perturbations are caused by $\dC(x, t)$ and $\db(x, t)$ with $t<t_\ast$. There is a time delay when a perturbation at the ice base is visible at the surface in $h$ but in $u$ it is observed immediately. As in the steady state, the same effect on $\du$ and $\dh$ can be achieved by different $\dC$ and $\db$. 
The higher the frequency of the perturbation is the more it is damped at the surface, but perturbations of low frequency reach the upper surface.

The numerical results in \citep{CGPL19a} confirm the conclusions here and are in good agreement with the analytical solutions.

%The SSA equations are derived from the FS equations in ~\ref{sec:SSA} following \citep{GreveBlatterBok, SSA}. 

\appendix

%!TEX root = ./iceinvanal.tex

\section{Derivation of the adjoint equations}\label{sec:appA}

\subsection{Adjoint viscosity and friction in SSA}\label{sec:adjvisc}

The adjoint viscosity $\tfateta(\fatu)$ in SSA in \eqref{eq:FSviscadj} is derived as follows. The SSA viscosity for $\fatu$ and $\fatu+\dfatu$ is 
\begin{equation}
\begin{array}{ll}\label{eq:adjviscSSA}
  % \eta(\fatu)&=&\frac{1}{2}A^{-1/n}\heta^\nu,\\
  % \heta&=&u_x^2+v_y^2+\frac{1}{4}(u_{1y}+u_{2x})^2+u_{1x}u_{2y}=\frac{1}{2}\fatB(\fatu):\fatD(\fatu),\\
  \eta(\fatu+\dfatu)\\
\approx \eta(\fatu)\left(1+\frac{1-n}{2n}\frac{(2u_{1x}+u_{2y})\du_{1x}+\frac{1}{2}(u_{1y}+u_{2x})\du_{2x}+(2u_{2y}+u_{1x})\du_{2y}+\frac{1}{2}(u_{1y}+u_{2x})\du_{1y}}{\heta}\right).
   % &=&\eta(\fatu)\left(1+\frac{1-n}{2n\heta}\fatB(\fatu):\fatD(\dfatu)\right)=\eta(\fatu)(1+\varrho(\fatu, \dfatu)).
\end{array}
\end{equation}
Determine $\calB(\fatu)$ such that
\[
   \varrho(\fatu, \dfatu)\fatB(\fatu)=\calB(\fatu)\star \fatB(\dfatu).
\]
First note that
\[
\begin{array}{ll}\label{eq:adjviscSSArel}
    \fatB(\fatu):\fatD(\dfatu)=(\fatD(\fatu)+\nabla\cdot\fatu\fatI):\fatD(\dfatu)=\fatD(\fatu):\fatD(\dfatu)+(\nabla\cdot\fatu)(\nabla\cdot\dfatu)\\
    =\fatD(\fatu):(\fatB(\dfatu)-\nabla\cdot\dfatu\fatI)+(\nabla\cdot\fatu)(\nabla\cdot\dfatu)=\fatD(\fatu):\fatB(\dfatu).
\end{array}
\]
Then use the $\star$ operator in \eqref{eq:stardef} to define $\calB$ 
\[
\begin{array}{rll}\label{eq:adjviscSSAstar}
  \frac{1-n}{2n\heta}\sum_{kl}B_{kl}(\fatu)D_{kl}(\dfatu) B_{ij}(\fatu)&=\frac{1-n}{2n\heta}\sum_{kl}D_{kl}(\fatu)B_{kl}(\dfatu) B_{ij}(\fatu)\\
  &=\sum_{kl}\calB_{ijkl}(\fatu)D_{kl}(\dfatu)=(\calB\star D)_{ij}.
\end{array}
\]
Thus, let
\[
   \calB_{ijkl}=\frac{1-n}{2n\heta} B_{ij}(\fatu)D_{kl}(\fatu),\quad \tilde{\eta}_{ijkl}(\fatu)=\eta(\fatu)(\calI_{ijkl}+\calB_{ijkl}(\fatu)),
\]
or in tensor form
\begin{equation}\label{eq:SSABdef}
   \calB=\frac{1-n}{n\fatB(\fatu):\fatD(\fatu)}\fatB(\fatu)\otimes \fatD(\fatu), \quad \tfateta(\fatu)=\eta(\fatu)\left(\calI+\calB\right).
\end{equation}
Replacing $\fatB$ in \eqref{eq:SSABdef} by $\fatD$ we obtain the adjoint FS viscosity in \eqref{eq:FSviscadj}. 

The adjoint friction in SSA in $\omega$ and at $\gamma_g$ in \eqref{eq:SSAadj3D} with a Weertman law is derived as in the adjoint FS equations \eqref{eq:FSadj}
and \eqref{eq:FSviscadj}. 
Then in $\omega$ with $\fatxi=\fatu, \fatzeta=\fatv, c=C, \fatF=\fatF_\omega$,
and at $\gamma_g$ with $\fatxi=\fatt\cdot\fatu, \fatzeta=\fatt\cdot\fatv, c=C_\gamma, f=f_\gamma, \fatF=F_\gamma,$ we arrive at the adjoint friction term $cf(\fatxi)\left(\fatI+\fatF(\fatxi)\right)\fatzeta$ where
\begin{equation}\label{eq:SSAFdef}
     \fatF(\fatxi)=\frac{m-1}{\fatxi\cdot\fatxi}\fatxi\otimes\fatxi.
\end{equation}

\subsection{Adjoint equations in SSA}\label{sec:appAdjointSSA}

The Lagrangian for the SSA equations is with the adjoint variables $\psi, \fatv, q$
\begin{equation}
\begin{array}{rll}\label{eq:SSALag}
   &\calL(\fatu, h;\fatv, \psi; b, C_\gamma, C)=\int_0^T\int_{\omega} F(\fatu, h)+\psi(h_t+\nabla\cdot(\fatu H)-a)\, \d\fatx\,\d t\\
       &+\int_0^T\int_{\omega} \fatv\cdot\nabla\cdot(2 H \eta\fatB(\fatu))-Cf(\fatu)\fatv\cdot\fatu-\rho g H\fatv\cdot\nabla h\,\, \d\fatx\,\d t\\
   &=\int_0^T\int_{\omega} F(\fatu, h)+\psi(h_t+\nabla\cdot(\fatu H)-a)\, \d\fatx\,\d t\\
   &+\int_0^T\int_{\omega} -2H\eta(\fatu)(\fatD(\fatv):\fatD(\fatu) +\nabla\cdot\fatu\nabla\cdot\fatv)\\
   &-Cf(\fatu)\fatv\cdot\fatu-\rho g H\fatv\cdot\nabla h \,\, \d\fatx\,\d t 
    -\int_0^T\int_{\gamma_g} C_\gamma f_\gamma(\fatt\cdot\fatu)\fatt\cdot\fatu\,\fatt\cdot\fatv\, \d s\,\d t
\end{array}
\end{equation}
after partial integration and using the boundary conditions.
The perturbed SSA Lagrangian is split into the unperturbed Lagrangian and three integrals
\begin{equation}
\begin{array}{lll}\label{eq:SSALagpert}
   \calL(\fatu+\dfatu, h+\dh; \fatv+\dfatv, \psi+\dpsi; b+\db, C_\gamma+\dC_\gamma, C+\dC)\\
    =\int_0^T\int_\omega F(\fatu+\dfatu, h+\dh)\\
    +\int_0^T\int_\omega (\psi+\dpsi)(h_t+\dh_t+\nabla\cdot\left((\fatu+\dfatu)(H+\dH)\right)-a)\, \d\fatx\,\d t\\
       +\int_0^T\int_\omega -2 (H+\dH) \eta(\fatu+\dfatu)\fatD(\fatv+\dfatv):\fatB(\fatu+\dfatu)\\
       -(C+\dC)f(\fatu+\dfatu)(\fatu+\dfatu)\cdot(\fatv+\dfatv)\\
    -\rho g(H+\dH)\nabla(h+\dh)\cdot(\fatv+\dfatv)\,\, \d\fatx\,\d t\\
   -\int_0^T\int_{\gamma_g} (C_\gamma+\dC_\gamma) f_\gamma(\fatt\cdot(\fatu+\dfatu))\fatt\cdot(\fatu+\dfatu)\,\fatt\cdot(\fatv+\dfatv)\, \d s\,\d t\\
   =\calL(\fatu, h; \fatv, \psi; b, C_\gamma, C)+I_1+I_2+I_3.
\end{array}
\end{equation}
The perturbation in $\calL$ is
\begin{equation}\label{eq:SSAdL}
    \delta\calL=I_1+I_2+I_3.
\end{equation}
Terms of order two or more in $\delta\calL$ are neglected. Then the first term in $\delta\calL$ satisfies
\begin{equation}
\begin{array}{rll}\label{eq:SSAI1}
   I_1&=\int_0^T\int_\omega F(\fatu+\dfatu, h+\dh)-F(\fatu, h)\, \d\fatx\,\d t\\
   &=\int_0^T\int_\omega F_\fatu\dfatu+F_h\dh \, \d\fatx\,\d t.
\end{array}
\end{equation}
Using partial integration, Gauss' formula, and the initial and boundary conditions on $\fatu$ and $H$ and $\psi(\fatx,T)=0, \fatx\in\omega,$ and 
$\psi(\fatx, t)=0, \;\fatx\in\gamma_w,$ in the second integral we have 
\begin{equation}
\begin{array}{rll}\label{eq:SSAI2}
   I_2&=&\int_0^T\int_\omega \dpsi(h_t+\nabla\cdot(\fatu H)-a)\\
   &&+\psi(\dh_t+\nabla\cdot(\dfatu H)+\nabla\cdot(\fatu \dH))\, \d\fatx\,\d t\\
   &=&\int_0^T\int_\omega \dpsi(h_t+\nabla\cdot(\fatu H)-a)\, \d\fatx\,\d t\\
   &&+\int_0^T\int_\omega-\psi_t\dh -H\nabla\psi\cdot\dfatu-\nabla\psi\cdot\fatu\dH\, \d\fatx\,\d t.
\end{array}
\end{equation}
The first integral after the second equality vanishes since $h$ is a weak solution and $I_2$ is 
\begin{equation}
\begin{array}{rll}\label{eq:SSAI2a}
   I_2=\int_0^T\int_\omega-(\psi_t+\fatu\cdot\nabla\psi)\dh -H\nabla\psi\cdot\dfatu+\fatu\cdot\nabla\psi\db\, \d\fatx\,\d t.
\end{array}
\end{equation}

%The weak solution $\fatu$ satisfies for any $\fatv$
%\begin{equation}
%\begin{array}{lll}\label{eq:SSAweak}
%   &\int_0^T\int_\omega -2H\eta(\fatu)\fatD(\fatv):\fatB(\fatu)
%    -Cf(\fatu)\fatu\cdot\fatv-\rho g H\nabla h\cdot\fatv\, \d\fatx\,\d t\\
%    =&\int_0^T\int_{\gamma_w}C_\gamma f_\gamma(\fatt\cdot\fatu)\fatt\cdot\fatu\,\fatt\cdot\fatv\, \d s\,\d t
%\end{array}
%\end{equation}
Using the weak solution of \eqref{eq:SSAforw3D}, the adjoint viscosity \eqref{eq:SSAviscadj3D}, \eqref{eq:SSABdef}, 
the friction coefficient \eqref{eq:SSAFdef}, Gauss' formula, the boundary 
conditions, and neglecting the second order terms, the third and fourth integrals in \eqref{eq:SSALagpert} are
\begin{equation}
\begin{array}{rll}\label{eq:SSAI3}
   I_3=&I_{31}+I_{32},\\
   I_{31}=&\int_0^T\int_\omega -2(H+\dH)\eta(\fatu+\dfatu)\fatD(\fatv+\dfatv):\fatB(\fatu+\dfatu)\\
   &-(C+\dC)f(\fatu+\dfatu)(\fatu+\dfatu)\cdot(\fatv+\dfatv)\\
   &-\rho g(H+\dH)\nabla(h+\dh)\cdot(\fatv+\dfatv))\,\, \d\fatx\,\d t\\
   &-\int_0^T\int_\gamma (C_\gamma+\dC_\gamma) f_\gamma(\fatt\cdot(\fatu+\dfatu)) \fatt\cdot(\fatu+\dfatu)\,\fatt\cdot(\fatv+\dfatv)\, \d s\,\d t\\
   =&I_{311}+I_{312}-I_{313},
\end{array}
\end{equation}
where
\begin{equation}
\begin{array}{rll}\label{eq:SSAI3a}
   I_{311}=&\int_0^T\int_\omega -2H\fatD(\fatv):(\eta(\fatu+\dfatu)\fatB(\fatu+\dfatu))\\
          &+2H\fatD(\fatv):(\eta(\fatu)\fatB(\fatu))\, \d\fatx\,\d t \\
      =&\int_0^T\int_\omega -2H\fatD(\fatv):(\tfateta(\fatu)\star \fatB(\dfatu))\, \d\fatx\,\d t\\
   I_{312}=&\int_0^T\int_{\omega_g}-\dC f(\fatu)\fatu\cdot\fatv\, \d\fatx\,\d t\\
          &+\int_0^T\int_{\omega_g}-C (f(\fatu+\dfatu)\fatv\cdot(\fatu+\dfatu)-f(\fatu)\fatv\cdot\fatu)\, \d\fatx\,\d t\\
         =& \int_0^T\int_{\omega_g} -\dC f(\fatu)\fatu\cdot\fatv
           +Cf(\fatu)(\fatI+\fatF_\omega(\fatu))\dfatu\cdot\fatv\, \d\fatx\,\d t \\
   I_{313}= &\int_0^T\int_{\gamma_g}(C_\gamma+\dC_\gamma) (f_\gamma(\fatt\cdot(\fatu+\dfatu))\fatt\cdot\fatv\,\fatt\cdot(\fatu+\dfatu)\\
              & -f_\gamma(\fatt\cdot\fatu)\fatt\cdot\fatv\,\fatt\cdot\fatu)\, \d s\,\d t\\
         =& \int_0^T\int_{\gamma_g} (C_\gamma+\dC_\gamma)(f_\gamma(\fatt\cdot\fatu)\fatt\cdot\fatu\,\fatt\cdot\fatv\\
          & +C_\gamma f_\gamma(\fatt\cdot\fatu)(\fatI+\fatF_\gamma(\fatt\cdot\fatu))\fatt\cdot\dfatu\,\fatt\cdot\fatv\, \d s\,\d t \\
   I_{32}=&\int_0^T\int_\omega-\rho g H\nabla h\cdot\fatv-2\eta\fatD(\fatv):\fatB(\fatu)\dH\\
                 &-\rho g \nabla h\cdot\fatv\dH-\rho g H \fatv\cdot\nabla \dh\, \d\fatx\,\d t\\
           =&\int_0^T\int_\omega-\rho g H\nabla h\cdot\fatv-(2\eta\fatD(\fatv):\fatB(\fatu)+\rho g \nabla h\cdot\fatv)\dH\\
            &+\rho g \nabla\cdot(H \fatv)\dh\, \d\fatx\,\d t.
\end{array}
\end{equation}
Collecting all the terms in \eqref{eq:SSAI1}, \eqref{eq:SSAI2a}, and \eqref{eq:SSAI3}, the first variation of $\calL$ is
\begin{equation}
\begin{array}{rll}\label{eq:SSA1stvar}
   \delta\calL&=&I_1+I_2+I_3\\
   &=&\int_0^T\int_\omega F_\fatu\dfatu-2HD(\fatv):(\tfateta(\fatu)\star \fatB(\dfatu))- H\nabla\psi\cdot\dfatu\, \d\fatx\,\d t\\
    &&-\int_0^T\int_{\omega_g} C f(\fatu)(\fatI+\fatF_\omega(\fatu))\fatv\cdot\dfatu\, \d\fatx\,\d t\\
    &&-\int_0^T\int_{\gamma_g}C_\gamma f_\gamma(\fatt\cdot\fatu)(\fatI+\fatF_\gamma(\fatt\cdot\fatu))\fatt\cdot\fatv\,\fatt\cdot\dfatu\, \d s\,\d t\\
   &&-\int_0^T\int_{\gamma_g}\dC_\gamma f_\gamma(\fatt\cdot\fatu) \fatt\cdot\fatu\,\fatt\cdot\fatv\, \d s\,\d t\\
   &&+\int_0^T\int_\omega (F_h  -(\psi_t+\fatu\cdot\nabla\psi +2\eta\fatD(\fatv):\fatB(\fatu)\\
   &&-\rho g \nabla b\cdot\fatv+\rho g H\nabla\cdot\fatv))\dh\, \d\fatx\,\d t\\
   &&+\int_0^T\int_\omega-\dC f(\fatu)\fatv\cdot\fatu\\
    &&+(2\eta \fatD(\fatv):\fatB(\fatu)+\rho g\nabla h\cdot\fatv+\fatu\cdot\nabla\psi)\db\, \d\fatx\,\d t.
\end{array}
\end{equation}
The forward solution $(\fatu^*, p^*, h^*)$ and adjoint solution $(\fatv^*, q^*, \psi^*)$ satisfying 
\eqref{eq:SSAforw3D} and \eqref{eq:SSAadj3D} are inserted into \eqref{eq:SSALag} resulting in
\begin{equation}
\begin{array}{rll}\label{eq:SSAopt}
   &\calL(\fatu^*, p^*; \fatv^*, q^*; h^*, \psi^*; b, C_\gamma, C)=\int_0^T\int_\omega F(\fatu^*, h^*)\, \d\fatx\,\d t.
\end{array}
\end{equation}
Then \eqref{eq:SSA1stvar} yields the variation in $\calL$ in \eqref{eq:SSAopt} with respect to perturbations $\db, \dC_\gamma,$ and $\dC$ in $b, C_\gamma,$ and $C$
\begin{equation}
\begin{array}{rll}\label{eq:SSA1stvar2}
   \delta\calL&=&\int_0^T\int_\omega(2\eta \fatD(\fatv^*):\fatB(\fatu^*)+\rho g\nabla h^*\cdot\fatv^*+\fatu^*\cdot\nabla\psi^*)\db\, \d\fatx\,\d t\\
              &&-\int_0^T\int_{\gamma_g}\dC_\gamma f_\gamma(\fatt\cdot\fatu^*) \fatt\cdot\fatu^*\,\fatt\cdot\fatv^*\, \d s\,\d t\\
              &&-\int_0^T\int_\omega \, \dC f(\fatu^*)\fatv^*\cdot\fatu^*\, \d\fatx\,\d t.
\end{array}
\end{equation}

\subsection{Adjoint equations in FS}\label{sec:appAdjointFS}

The  FS Lagrangian is
\begin{equation}
\begin{array}{rll}\label{eq:FSLag}
   &\calL(\fatu, p, h; \fatv, q, \psi; C)=\int_0^T\int_{\Gamma_s} F(\fatu, h)+\psi(h_t+\fath\cdot\fatu-a)\, dx\,\d t\\
       &+\int_0^T\int_\omega\int_b^h -\fatv\cdot(\nabla\cdot\fatsigma(\fatu, p))-q\nabla\cdot\fatu-\rho \fatg\cdot\fatv\,\, \d\fatx\,\d t\\
   &=\int_0^T\int_{\Gamma_s} F(\fatu, h)+\psi(h_t+\fath\cdot\fatu-a)\, \d\fatx\,\d t\\
   &+\int_0^T\int_\omega\int_b^h 2\eta(\fatu)\fatD(\fatv):\fatD(\fatu) -p\nabla\cdot\fatv-q\nabla\cdot \fatu-\rho \fatg\cdot\fatv \,\, \d\fatx\,\d t \\
   &   +\int_0^T\int_{\Gamma_b} C f(\fatT\fatu)\fatT\fatu\cdot\fatT\fatv\, \d\fatx\,\d t.
\end{array}
\end{equation}
In the same manner as in \eqref{eq:SSALagpert}, the perturbed FS Lagrangian is
\begin{equation}
\begin{array}{lll}\label{eq:FSLagpert}
   \calL(\fatu+\dfatu, p+\ddp; \fatv+\dfatv, q+\dq; h+\dh, \psi+\dpsi; C+\dC)\\
   =\calL(\fatu, p, h; \fatv, q, \psi;  C)+I_1+I_2+I_3.
\end{array}
\end{equation}
Terms of order two or more in $\dfatu, \dfatv, \dh$ are neglected. The first integral $I_1$ in \eqref{eq:FSLagpert} is
\begin{equation}
\begin{array}{rll}\label{eq:FSI1}
   I_1&=\int_0^T\int_{\Gamma_s} F(\fatu(\fatx,h+\dh,t)+\dfatu, h+\dh)-F(\fatu(\fatx,h,t), h)\, \d\fatx\,\d t\\
   &=\int_0^T\int_{\Gamma_s} F_\fatu(\dfatu+\fatu_z\dh)+F_h\dh \, \d\fatx\,\d t.
\end{array}
\end{equation}
Partial integration, the conditions $\psi(\fatx,T)=0$ and $\psi(\fatx, t)=0$ at $\Gamma_s$, and the fact that $h$ is a weak solution
simplify the second integral 
\begin{equation}
\begin{array}{rll}\label{eq:FSI2}
   I_2&=&\int_0^T\int_{\Gamma_s} \dpsi(h_t+\fath\cdot\fatu-a)\\
   &&+\psi(\dh_t+\fatu\cdot\dfath+\fatu_z\cdot\fath\dh+\fath\cdot\dfatu)\, \d\fatx\,\d t\\
   &=&\int_0^T\int_{\Gamma_s} \dpsi(h_t+\fath\cdot\fatu-a)\, \d\fatx\,\d t\\
   &&+\int_0^T\int_{\Gamma_s}(-\psi_t-\nabla\cdot(\fatu\psi)+\fath\cdot\fatu_z\psi)\dh+\fath\cdot\dfatu\psi\, \d\fatx\,\d t.
\end{array}
\end{equation}

Define $\Xi,\xi,$ and $\Upsilon$ to be 
\begin{equation}
\begin{array}{rll}\label{eq:FSdef}
   \Theta(\fatu, p; \fatv, q; C)&=&2\eta(\fatu)\fatD(\fatv):\fatD(\fatu)
    -p\nabla\cdot\fatv - q\nabla\cdot\fatu-\rho\fatg\cdot\fatv,\\
   \theta(\fatu;\fatv;C)&=&C f(\fatT\fatu)\fatT\fatu\cdot\fatT\fatv,\\
   \Upsilon(\fatu, p; \fatv, q)&=&-\fatv\cdot(\nabla\cdot\fatsigma(\fatu, p))
   -q\nabla\cdot\fatu-\rho\fatg\cdot\fatv.
\end{array}
\end{equation}
Then a weak solution, $(\fatu, p)$, for any $(\fatv, q)$ satisfying the boundary conditions, fulfills 
\begin{equation}
\label{eq:FSweak}
   \int_0^T\int_\omega\int_b^{h} \Theta(\fatu, p; \fatv, q; C)\, \d\fatx\,\d t-\int_0^T\int_{\Gamma_b}\theta(\fatu;\fatv;C)\, \d\fatx\,\d t=0.
\end{equation}
The third integral in \eqref{eq:FSLagpert} is
\begin{equation}
\begin{array}{rll}\label{eq:FSI3}
   I_3=&I_{31}+I_{32},\\
   I_{31}=&\int_0^T\int_\omega\int_b^{h} \Theta(\fatu+\dfatu, p+\ddp; \fatv+\dfatv, q+\dq; C+\dC)\, \d\fatx\,\d t\\
         &-\int_0^T\int_{\Gamma_b}\theta(\fatu+\dfatu;\fatv+\dfatv;C+\dC)\, \d\fatx\,\d t,\\
  I_{32}=&\int_0^T\int_\omega\int_h^{h+\dh} \Upsilon(\fatu, p; \fatv, q)\, \d\fatx\,\d t.
\end{array}
\end{equation}
The integral $I_{31}$ is expanded as in \eqref{eq:SSAI3} and \eqref{eq:SSAI3a} or \citep{Petra12} using the weak solution, Gauss' formula, 
and the definitions of the adjoint viscosity and adjoint friction coefficient in Sect.~\ref{sec:adjvisc}.
When $b<z<h$ we have $\Upsilon(\fatu, p; \fatv, q)=0$. 
If $\Upsilon$ is extended smoothly in the positive $z$-direction from $z=h$, then with  $z\in[h, h+\dh]$ for some constant $c>0$ we have $|\Upsilon|\le c\dh$. 
Therefore,
\[
   |\int_h^{h+\dh(x,t)} \Upsilon(\fatu, p; \fatv, q)\, dz|\le \int_h^{h+\dh(x,t)} \sup |\Upsilon|\, dz\le c|\dh(x,t)^2|,
\]
and the bound on $I_{32}$ in \eqref{eq:FSI3} is 
\begin{equation}
\label{eq:FSI32est}
   |I_{32}|\le ct|\omega| \max|\dh(x, t)|^2,
\end{equation}
where $|\omega|$ is the area of $\omega$.
This term is a second variation in $\dh$ which is neglected and $I_3=I_{31}$.

%\begin{equation}
%\begin{array}{rll}\label{eq:FSI4}
%   \int_0^T\int_\omega -(C+\dC)f(\|\fatT(\fatu+\dfatu)\|) \fatT(\fatu+\dfatu)\cdot\fatT(\fatv+\dfatv)\, dx\,\d t
%\end{array}
%\end{equation}

The first variation of $\calL$ is then
\begin{equation}
\begin{array}{rll}\label{eq:FS1stvar}
   \delta\calL&=&I_1+I_2+I_3\\
   &=&\int_0^T\int_{\Gamma_s} (F_\fatu + \psi\fath)\cdot\dfatu\, \d\fatx\,\d t\\
   &&+\int_0^T\int_{\Gamma_s} (F_h +F_\fatu \fatu_z -(\psi_t+\nabla\cdot(\fatu\psi) -\fath\cdot\fatu \psi))\dh\, \d\fatx\,\d t\\
   &&+\int_0^T\int_\omega\int_b^{h} 2\fatD(\fatv):(\tfateta(\fatu)\star \fatD(\dfatu))-\ddp \nabla\cdot\fatv - q\nabla\cdot\dfatu\, \d\fatx\,\d t\\
   &&+\int_0^T\int_{\Gamma_b}C f(\fatT\fatu)(\fatI+\fatF_b(\fatu))\fatT\fatv\cdot\fatT\dfatu\, \d\fatx\,\d t\\
   &&+\int_0^T\int_{\Gamma_b} \dC f(\fatT\fatu)\, \fatT\fatu\cdot\fatT\fatv\, \d\fatx\,\d t.
\end{array}
\end{equation}
With the forward solution $(\fatu^*, p^*, h^*)$ and the adjoint solution $(\fatv^*, q^*, \psi^*)$ satisfying 
\eqref{eq:FSforw} and \eqref{eq:FSadj}, the first variation with respect to perturbations $\dC$ in $C$ is (cf. \eqref{eq:SSA1stvar2})
\begin{equation}
\label{eq:FS1stvar2}
   \delta\calL=\int_0^T\int_{\Gamma_b} f(\fatT\fatu^*)\, \fatT\fatu^*\cdot\fatT\fatv^*\, \dC \, \d\fatx\,\d t.
\end{equation}

%!TEX root = ./iceinvanal.tex

% \section{Analytical solutions}\label{sec:appB}

% {\color{red} After simplifications justified by the numerical experiments in \cref{sec:results}, analytical solutions to the SSA forward and adjoint equations are derived in 2D.}

%%%%%%%%%%%%%%%%%%%%%%%%%%%%%%%%%%%%%%%%%%%%%%%%%%%%%%%%%%%%%%%%%%%%%%%%%%%%%%%%%%%
\section{Simplified SSA equations}\label{sec:SSAanalyt}

The forward and adjoint SSA equations in \eqref{eq:SSAforwsimp} and \eqref{eq:SSAadjsimp} are solved analytically.
The conclusion from the thickness equation in \eqref{eq:SSAforwsimp} is that 
\begin{equation}
\label{eq:SSAuHsol}
    u(x)H(x)=u(0)H(0)+ax=ax, 
\end{equation}
since $u(0)=0$.
Solve the second equation in \eqref{eq:SSAforwsimp} for $u$ on the bedrock with $x\leq x_{GL}$ and insert into \eqref{eq:SSAuHsol} using the assumptions for $x>0$ that $b_x\ll H_x$ and $h_x\approx H_x$ to have 
\begin{equation}\label{eq:SSAintH}
  \frac{\rho g}{C}H^{m+1}H_x =\frac{\rho g}{C(m+2)}(H^{m+2})_x = -(ax)^m.
\end{equation}
The equation for $H^{m+2}$ for $x\leq x_{GL}$ is integrated from $x$ to $x_{GL}$ such that
\begin{equation}
\begin{array}{rll}\label{eq:SSAHsol}
   H(x)&=\displaystyle{\left(H^{m+2}_{GL}+\frac{m+2}{m+1}\frac{C a^m}{\rho g}(x_{GL}^{m+1}-x^{m+1})\right)^{\frac{1}{m+2}}},\\
   u(x)&=\displaystyle{\frac{ax}{H}}, \quad H_x=\displaystyle-\frac{Ca^m}{\rho g}\frac{x^m}{H^{m+1}}.
\end{array}
\end{equation}
For the floating ice at $x>x_{GL}$, $\rho g H h_x=0$ implying that $h_x=0$ and $H_x=0$. Hence, $H(x)=H_{GL}$. 
The velocity increases linearly beyond the grounding line
\begin{equation}\label{eq:SSAusol}
    u(x)={ax}/{H(x)}={ax}/{H_{GL}},\; x>x_{GL}.
\end{equation}
By including the viscosity term in \eqref{eq:SSAforw} and assuming that $H(x)$ is linear in $x$, a more accurate formula is obtained for $u(x)$ on the floating ice in
(6.77) of \citep{GreveBlatterBok}.

%%%%%%%%%%%%%%%%%%%%%%%%%%%%%%%%%%%%%%%%%%%%%%%%%%%%%%%%%%%%%%%%%%%%%%%%%%%%%%%%%%%
\section{Jumps in $\psi$ and $v$ in SSA}\label{sec:SSAjumps}

Multiply the first equation in \eqref{eq:SSAadjsimp} by $H$ and the second equation by $u$ to eliminate $\psi_x$. We get
\begin{equation}\label{eq:SSAadjsimp3} 
  -Cmu^mv-\rho g H^2 v_x=HF_h-uF_u.
\end{equation}
Use the expression for $u$ and $H_x$ in \eqref{eq:SSAHsol}. Then
\begin{equation}\label{eq:SSAadjsimp4}
    \rho gH(mH_xv-Hv_x)=HF_h-uF_u, 
\end{equation}
or equivalently
\begin{equation}\label{eq:SSAadjsimp_v}
  \left(\frac{v}{H^m} \right)_x = -\frac{1}{\rho gH^{m+2}}(HF_h-uF_u). 
\end{equation}

The solutions $\psi(x)$ and $v(x)$ of the adjoint SSA equation \eqref{eq:SSAadj} have jumps at the observation point $x_\ast$.
For $x$ close to $x_{*}$ in a short interval $[x_\ast^-, x_\ast^+]$ with $x_\ast^-<x_\ast<x_\ast^+$, integrate \eqref{eq:SSAadjsimp_v} to receive
\begin{equation}\label{eq:vjump}
  \int_{x_{*}^-}^{x_{*}^+}  \left(\frac{v}{H^m}\right)_x\, \d x= -\int_{x_{*}^-}^{x_{*}^+} \frac{HF_h-uF_u}{\rho gH^{m+2}}\, \d x.
\end{equation}
Since $H$ is continuous and $u$ and $v$ are bounded, when $x_{*}^-\rightarrow x_{*}^+$, then
% the left integral in \eqref{eq:vjump} is
% \begin{equation}\label{eq:vjump left}
%   \int_{x_{*}^-}^{x_{*}^+}  \left(\frac{v}{H^m}\right)_x\, \d x= \frac{v(x_{*}^+)-v(x_{*}^-)}{H_\ast^m},
% \end{equation}
% and the right integral is
% \begin{equation}\label{eq:vjump right}
%   -\int_{x_{*}^-}^{x_{*}^+} \frac{HF_h-uF_u}{\rho gH^{m+2}}\, \d x= -\frac{H_\ast\int_{x_{*}^-}^{x_{*}^+}F_h\, \d x - u_\ast \int_{x_{*}^-}^{x_{*}^+}F_u \, \d x}{\rho gH_\ast^{m+2}}.
% \end{equation}
% Therefore
\begin{equation}\label{eq:vjumpInt}
  v(x_{*}^+)-v(x_{*}^-) = -\frac{1}{\rho gH_\ast^{2}} \left(H_\ast\int_{x_{*}^-}^{x_{*}^+}F_h\, \d x - u_\ast \int_{x_{*}^-}^{x_{*}^+}F_u \, \d x\right).
\end{equation}
A similar relation for $\psi$ can be derived
\begin{equation}\label{eq:vjumpInt2}
  \psi(x_{*}^+)-\psi(x_{*}^-) =\frac{1}{H_\ast} \int_{x_{*}^-}^{x_{*}^+}F_u \, \d x .
\end{equation}
With $F_u=0$ and $F_h=0$ for $x<x_\ast$ and  $v(0)=\psi_x(0)=0$, we find that 
\begin{equation}
\label{eq:SSAvpsileft}
    v(x)=\psi_x(x)=0,\quad \psi(x) = \psi(x_*^-),\; 0\leq x<x_\ast.
\end{equation}

If $F(u,h)=u\delta(x-x_\ast)$, then by \eqref{eq:vjumpInt} and \eqref{eq:vjumpInt2}
\begin{equation}\label{eq:jumpUcond}
  v(x_{*}^+)=\frac{u_\ast}{\rho g H_\ast^2},\quad \psi(x_{*}^+)-\psi(x_{*}^-)=\frac{1}{H_\ast},
\end{equation}
and if $F(u,h)=h\delta(x-x_\ast)$, then
\begin{equation}\label{eq:jumpHcond}
  v(x_{*}^+)=-\frac{1}{\rho g H_\ast},\quad, \psi(x_{*}^+)-\psi(x_{*}^-)=0.
\end{equation}

%%%%%%%%%%%%%%%%%%%%%%%%%%%%%%%%%%%%%%%%%%%%%%%%%%%%%%%%%%%%%%%%%%%%%%%%%%%%%%%%%%%
\section{Analytical solutions in SSA}\label{sec:analyticalAdjointSSA}

%%%% v %%%%
By Sect.~\ref{sec:SSAjumps}, $v(x)=0$  for $0\leq x<x_{\ast}$.
Use equations in \eqref{eq:SSAadjsimp} with $H_x$ in \eqref{eq:SSAHsol} for $x_\ast<x\leq x_{GL}$ to have 
\[
 \frac{v_x}{v}=-\frac{axCmu^{m-1}}{\rho g H^3}=-\frac{Cmu^{m}}{\rho g H^2}=\frac{mH_x}{H}.
\]
% Hence, $v$ is
% \begin{equation}\label{eq:SSAvsol}
%     v(x)=C_v H(x)^m, \; x_\ast< x\leq x_{GL}.
% \end{equation}
Let $\mathcal{H}(x-x_\ast)=\int_{-\infty}^{x-x_\ast} \delta(s)\, ds$ be the Heaviside step function at $x_\ast$. Then
\begin{equation}\label{eq:SSAvsol}
  v(x) = C_vH(x)^m\mathcal{H}(x-x_\ast),\quad 0\leq x\leq x_{GL}.
\end{equation}
To satisfy the jump condition in \eqref{eq:jumpUcond} and \eqref{eq:jumpHcond}, the constant $C_v$ is  
\begin{equation}\label{eq:SSAvsolCoeff}
  C_v = 
  \begin{cases}
  \begin{array}{rll}
    \displaystyle{\frac{ax_\ast}{\rho g H_\ast^{m+3}}},\quad& F(u,h)=u\delta(x-x_\ast),\\
    \displaystyle{-\frac{1}{\rho g H_\ast^{m+1}}},\quad&   F(u,h)=h\delta(x-x_\ast).
  \end{array}
  \end{cases}
\end{equation}

%%%% psi %%%%
Combine \eqref{eq:SSAvsol} with the relation $\psi_x=(F_u-Cm u^{m-1}v)/H$ and integrate from $x$ to $x_{GL}$ to obtain
\begin{equation}\label{eq:SSApsisol}
  \psi(x)= C_va^{m-1}C\left(x_{GL}^m - x^m\right), \; x_\ast< x\leq x_{GL}.
\end{equation}
With the jump condition in \eqref{eq:jumpUcond} and \eqref{eq:jumpHcond}, $\psi(x)$ at $0\leq x<x_\ast$ is
\begin{equation}\label{eq:SSApsisol2}
  \psi(x)=
  \begin{cases}
  \begin{array}{rlll}
    &\displaystyle{-\frac{1}{H_\ast}+\frac{C a^m x_\ast}{\rho g H_\ast^{m+3}}\left(x_{GL}^m - x_\ast^m\right)},\quad& F(u,h)=u\delta(x-x_\ast),\\
    &\displaystyle{-\frac{C a^{m-1}}{\rho g H_\ast^{m+1}}(x_{GL}^m - x_\ast^m)},\quad&  F(u,h)=h\delta(x-x_\ast).
  \end{array}
  \end{cases}
\end{equation}

%%%% weight C %%%%
The weight for $\delta C$ in the functional $\delta\calL$ in \eqref{eq:SSAgrad} is non-zero for $x_\ast< x\leq x_{GL}$
\begin{equation}\label{eq:SSAvu}
   -v u^m=-C_v(ax)^m.
\end{equation}

%%%% weight b %%%%
Use \eqref{eq:SSAvsol} and \eqref{eq:SSAadjsimp} in \eqref{eq:SSAgrad} to determine the weight for $\db$ in $\dL$,
\begin{equation}\label{eq:SSAbweight}
    \begin{array}{llll}

  &\psi_x u+v_x\eta u_x+v \rho gh_x
  = \rho g (Hv)_x{+F_h}\\
  &=C_v\rho g H^m\left[(m+1)H_x\mathcal{H}(x-x_\ast) +H\delta(x-x_\ast)\right]+F_h.
  % =&
  %  \begin{cases}
  %   \begin{array}{rlll}
  %     &0,\quad& 0\leq x < x_\ast\\
  %     & v(x_\ast^+) ???,\quad& {\color{red} x=x_\ast},\\
  %     & \displaystyle{ -(m+1)C_vC \frac{(ax)^m}{H}},\quad& x_\ast<x\leq x_{GL}.
  %   \end{array}
  %  \end{cases}
    \end{array}
\end{equation}

% %%%%%%%%%%%%%%%%%%%%%%%%%%%%%%%%%%%%%%%%%%%%%%%%%%%%%%%%%%%%%%%%%%%%%%%%%%%%%%%%%%%%%%%%%%%%%%%%%%%%%%%%%%%%%%%%%%%%%%%%%%%%%%%%%%%%%%%%%%%%%%%%%%%%%
% \begin{acknowledgements}
\section*{Acknowledgement}

This work was supported by Nina Kirchner's Formas grant 2017-00665. 
Lina von Sydow read a draft of the paper and helped us improve the presentation with her comments.
% \end{acknowledgements}
%%%%%%%%%%%%%%%%%%%%%%%%%%%%%%%%%%%%%%%%%%%%%%%%%%%%%%%%%%%%%%%%%%%%%%%%%%%%%%%%%%%%%%%%%%%%%%%%%%%%%%%%%%%%%%%%%%%%%%%%%%%%%%%%%%%%%%%%%%%%%%%%%%%%%

\bibliographystyle{MG}       
\bibliography{icebib}

\end{document}